\RequirePackage{fix-cm}
\documentclass[twocolumn,epjc3]{svjour3}  
\smartqed  
\RequirePackage{graphicx}
\RequirePackage[colorlinks,citecolor=blue,urlcolor=blue,linkcolor=blue]{hyperref}
\journalname{Eur. Phys. J. C}

\newcommand{\ZvvHH}{\ensuremath{\PZ(\to\nunubar)\PH\PH}\xspace}
\newcommand{\gHHH}{\ensuremath{g_{\PH\PH\PH}}\xspace}
\newcommand{\gHHWW}{\ensuremath{g_{\PH\PH\PW\PW}}\xspace}
\newcommand{\HHWW}{\ensuremath{\PH\PH\PW\PW}\xspace}
\newcommand{\HHH}{\ensuremath{\PH\PH\PH}\xspace}
\newcommand{\gHHHSM}{\ensuremath{g_{\PH\PH\PH}^{\mathrm{SM}}}\xspace}
\newcommand{\gHHWWSM}{\ensuremath{g_{\PH\PH\PW\PW}^{\mathrm{SM}}}\xspace}

\newcommand{\deltachisquared}{\ensuremath{\Delta\chi^{2}}\xspace}
\newcommand{\Zhh}{\ensuremath{\PZ\PH\PH}\xspace}
\newcommand{\ZHH}{\ensuremath{\PZ\PH\PH}\xspace}
\newcommand{\HHveve}{\ensuremath{\PH\PH\nuenuebar}\xspace} 
\newcommand{\HHvv}{\ensuremath{\PH\PH\nunubar}\xspace} 

\newcommand{\Hvv}{\ensuremath{\PH\nunubar}\xspace}
\newcommand{\kappaHHWW}{\ensuremath{\kappa_{\PH\PH\PW\PW}}\xspace}
\newcommand{\kappaHHH}{\ensuremath{\kappa_{\PH\PH\PH}}\xspace}
\newcommand{\WW} {\ensuremath{\PW\PWst}\xspace} 
\usepackage{CLICdp_definitions}
\usepackage{heppennames2,siunitx,amssymb,booktabs,multirow}
\usepackage{lineno}
\begin{document}

\title{Double Higgs boson production and Higgs self-coupling extraction at CLIC
}


\author{Philipp Roloff\thanksref{addr1}
  \and
  Ulrike Schnoor\thanksref{e1,addr1}
  \and Rosa Simoniello\thanksref{e2,addr1}
  \and Boruo Xu\thanksref{addr2}
}

\thankstext{e1}{e-mail: ulrike.schnoor@cern.ch}
\thankstext{e2}{now at: Johannes Gutenberg University of Mainz, Germany}


\institute{CERN, Geneva, Switzerland \label{addr1}
           \and
           University of Cambridge, United Kingdom \label{addr2}
}

\date{Received: date / Accepted: date}

\maketitle

\begin{abstract}
  The Compact Linear Collider (CLIC) is a
future electron-positron collider that will allow measurements of the
trilinear Higgs self-coupling in double Higgs boson events produced at
its high-energy stages with collision energies from
$\sqrt{s}$\,=\,1.4\,TeV to 3\,TeV.
  The sensitivity to the  Higgs self-coupling is driven by the mea\-sure\-ments of the  cross section and the  invariant mass distribution of the Higgs-boson pair in the W-boson
fusion process, $\epem\to\HHvv$. It is enhanced by including the
  cross-section measurement of \Zhh production at 1.4\,TeV.
  The expected sensitivity of CLIC for Higgs pair production through 
W-boson fusion is studied for the decay channels \bbbb and \bbWW using full detector
simulation including all relevant backgrounds
at $\roots=\SI{1.4}{\TeV}$ with an integrated luminosity of  $\mathcal{L}$\,=\,2.5\,ab$^{-1}$ 
and at $\roots=\SI{3}{\TeV}$ with $\mathcal{L}$\,=\,5\,ab$^{-1}$.
Combining $\epem\to\HHvv$ and \Zhh cross-section
measurements at 1.4\,TeV with differential measurements in
$\epem\to\HHvv$ events at 3\,TeV, CLIC will be able to measure the trilinear
Higgs self-coupling with a relative uncertainty of $-8\,\%$ and $
+11\,\%$ at 68\,\% C.L., assuming the Standard Model.
In addition, prospects for simultaneous constraints on the trilinear
Higgs self-coupling and the Higgs-gauge coupling HHWW are derived based on the \HHvv measurement.
\end{abstract}

\section{Introduction}
The discovery of the Higgs boson \cite{Aad:2012tfa,Chatrchyan:2012ufa} has initiated an era of investigations of its properties and of the nature of the mechanism that breaks the electroweak symmetry.
Besides its mass and width, the properties of interest include the
couplings of the Higgs boson to other Standard Model (SM) and
hypothetical non-SM particles as well as the coupling to itself.
While the couplings to other SM particles illustrate the way these
particles obtain masses in the Higgs mechanism, the self-coupling
parameter determines the shape of the Higgs potential which has
implications for the vacuum metastability, the hierarchy problem, as well as the electroweak phase transition and baryogenesis.
In the Standard Model, the Higgs potential for the Higgs field $\phi$ is described by
\begin{equation}
  \label{eq:higgspotential}
    V(\phi) = \mu^{2}  \phi^{\dagger} \phi + \frac{\lambda^2}{2}
    (\phi^{\dagger} \phi)^{2},
\end{equation}
where $\mu$ is proportional to the Higgs boson mass and $\lambda$ is the
Higgs self-coupling.
This implies a fixed relation $m_H^2 = \lambda v$ between the mass
and the self-coupling, with the vacuum expectation value $v$.
In the interaction Lagrangian, this potential leads to a trilinear
self-coupling \gHHH which is proportional to $\lambda$.

A deviation of the Higgs potential from the SM would directly point to new physics, for example in the context of baryogenesis:
Indeed, one of the conditions for electroweak baryogenesis is the presence of a strong first-order phase transition in the breaking of the electroweak symmetry in the early universe.
In order to modify the Higgs potential accordingly, at least one additional scalar needs to be introduced~\cite{Cohen:1993nk}.
This can be an additional scalar singlet~\cite{Davoudiasl:2004be} or doublet.
The latter is realised in two-Higgs doublet models (2HDM)~\cite{Gunion:425736,Branco:2011iw}, which introduce four additional scalars.
These models can lead to modifications of the Higgs self-coupling.
A sufficiently heavy neutral scalar can cause a resonance in the
invariant mass of the Higgs boson pair in the production of two
SM-like Higgs bosons.

Existing models including those discussed above with additional scalars as well as theories where the Higgs boson is composite predict differences of the Higgs self-couplings to the
SM value between a few and tens of percent~\cite{Gupta:2013zza}.
These estimates assume
the scenario that no additional states of the
electroweak symmetry breaking sector can be discovered at the LHC.
An overview of BSM theories modifying the Higgs self-coupling
is given in~\cite{HHwhitepaper}.

A measurement of the Higgs self-coupling with a precision of better than 50\,\%
will not be possible at the High-Luminosity Large Hadron Collider
(HL-LHC)~\cite{Cepeda:2019klc}.
More precise measurements are possible at high-energy linear
colliders, as they give direct access to double Higgs boson
production in a comparably clean environment.
Electron-positron colliders below a center-of-mass energy of $\sqrt{s}
\approx 500$\,GeV do not have access to double Higgs boson production.
They can only constrain the Higgs self-coupling
indirectly through its loop contributions to single Higgs boson production~\cite{McCullough:2013rea}.
The prospects for several proposed future options are discussed in~\cite{deBlas:2019rxi}.
The potential of the International Linear Collider (ILC) to measure
the Higgs self-coupling directly in double Higgs boson production
in association with a \PZ boson
 at
$\sqrt{s}=500$\,GeV and in the W-boson fusion double Higgs production channel at  1\,TeV is described
in~\cite{duerig,Tian:2013qmi,kurata}.

The Compact Linear  Collider (CLIC) is a mature option for a future
linear electron-positron collider~\cite{cdrvol2}, which will allow the precise determination of the properties of the Higgs boson well beyond the precision of the HL-LHC. 
A detailed investigation of the CLIC prospects for the Higgs couplings
to SM particles is given in~\cite{Abramowicz:2210491} and an update of
these results to a new luminosity and polarisation baseline scenario is provided in~\cite{Roloff:2645352,Robson:2020lhl}.
A preliminary study of the Higgs self-coupling measurement at CLIC,
based only on the measurement of the double Higgs boson production, has
been presented in~\cite{Abramowicz:2210491}. The analysis is updated
and extended in this paper, most importantly by exploiting
differential distributions in the analysis of $\epem \to \HHvv$ at 3\,TeV, by
illustrating the impact of $\epem \to \Zhh $ at 1.4\,TeV, and by
extracting \gHHH and \gHHWW in a joint fit.

Each energy stage at CLIC contributes to the indirect measurement of
the Higgs self-coupling in single Higgs boson production.
Combined with the HL-LHC standalone precision of 47\,\% in the
one-parameter fit, the CLIC run
at the collision energy of 380\,GeV will only improve this precision to
46\,\%~\cite{deBlas:2019rxi}.
With increasing statistics and energy,
the indirect limits in the one-parameter fit will be improved to
41\,\% after the energy stage at $\sqrt{s}$~=~1.5\,TeV and 35\,\% after the 3\,TeV energy stage. 
However, already at 1.5\,TeV,
the direct accessibility of double Higgs production
allows much more powerful, potentially model-independent constraints to be put on the Higgs
self-coupling
which by far exceed the
precision obtained in single Higgs
measurements~\cite{deBlas:2019rxi}. These measurements are the subject
of this paper.

The high-energy stages of CLIC with centre-of-mass energies of 1.5 and
3\,TeV provide the opportunity to  access directly the trilinear Higgs
self-coupling in double Higgs boson production.
In the present study, the earlier choice
of centre-of-mass energy for the second stage of
1.4\,TeV~\cite{CLIC:2016zwp} is used due to the availability of full
simulation event samples.
While we therefore base the following study on a run at 1.4\,TeV, the
prospects for 1.5\,TeV are expected be very similar.
The main channels are double Higgsstrahlung \Zhh production at
1.4\,TeV and double Higgs boson production via W-boson  fusion  at 1.4
and 3\,TeV.
Both are directly sensitive to the trilinear Higgs self-coupling \gHHH, while the latter is also sensitive to the quartic Higgs-gauge coupling \gHHWW.
This paper uses full detector simulation to study the  CLIC potential for extracting these couplings
from measurements of double Higgs boson production.

In a full Effective Field Theory approach,  other operators apart from the one
modifying the triple Higgs vertex can also contribute to the same
final state. As these operators are themselves constrained by other
measurements, e.g.~single Higgs boson production channels,  a global
fit approach as studied in~\cite{DiVita:2017vrr} is appropriate.
Results for CLIC are presented in~\cite[Sec. 2.2.1]{BSRYR}, showing
that the constraints from the global fit are very close to the ones
obtained in
the exclusive approach, due to the high precision measurements of
other processes at CLIC.
A detailed study of the impact of other operators was performed for
the ZHH channel at 500\,GeV~\cite{Barklow:2017awn}.

This paper investigates the prospects for extracting the trilinear Higgs self-coupling at CLIC in double Higgs boson production at the high-energy stages of CLIC.
It is structured as follows:
Sec.~\ref{sec:strategy} describes the strategy of the analysis and the various contributions to the sensitivity. In Sec.~\ref{sec:SimulationSamples}, the definition of the signal and background processes, as well as the simulation and reconstruction chain, are described.
The event selection procedures for the analyses at 1.4 and 3\,TeV for \HHvv $\to \bbbb\nunubar$ and $\bbWW\nunubar$ are explained in Sec.~\ref{sec:eventselection}.
This is followed by the results for the cross section measurement in Sec.~\ref{sec:ResultsFromXSec} and for the differential measurement giving the most stringent constraints in Sec.~\ref{sec:ResultsFromDifferential}. A summary is provided in Sec.~\ref{sec:summary}.

\section{Analysis strategy}\label{sec:strategy}
\begin{figure}
\includegraphics[width=\columnwidth]{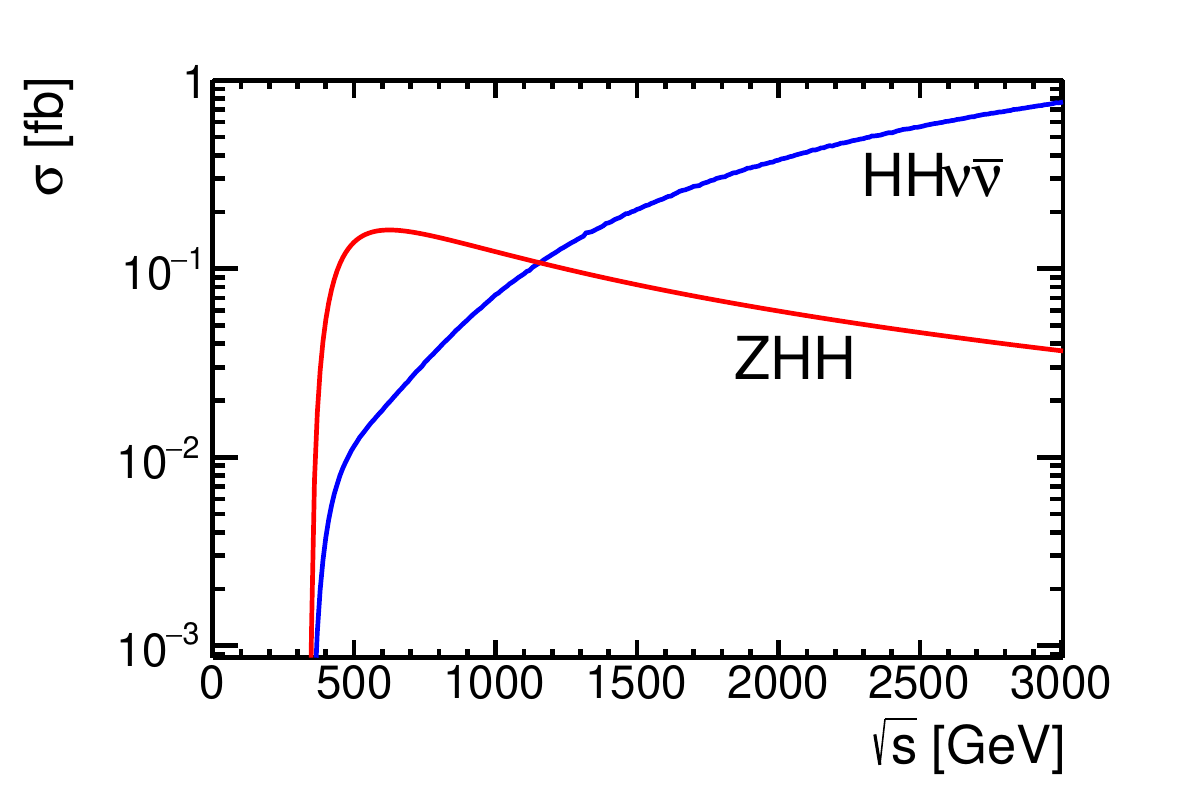}
\caption{Cross section as a function of centre-of-mass energy
for  $\epem\to \Zhh$ and $\epem\to \HHvv$ production for a Higgs boson mass
of $m_{\mathrm{H}}$~=~126 GeV. The values shown correspond to
unpolarised beams including initial state radiation but not including the effect
of beamstrahlung~\cite{Abramowicz:2210491}.}
\label{fig:xsec_energy}
\end{figure}

At CLIC, the Higgs self-coupling can be directly accessed through the measurement of double Higgs boson production.
Two main channels contribute: 
W-boson fusion (WBF) double Higgs boson production (dominant part of $\epem\to \HHvv$) and the  double Higgsstrahlung process  ($\epem\to \Zhh$).
The other process of vector boson fusion, namely \PZ-boson fusion
($\epem\to \PH\PH \epem$), has a one order of magnitude smaller cross section and is therefore not considered here.
The dependence of the cross section on the centre-of-mass energy 
obtained with
\textsc{Whizard~1.95}~\cite{Kilian:2007gr,Moretti:2001zz} is shown in
Fig.~\ref{fig:xsec_energy}.
This illustrates that
the highest cross section of \Zhh production among the forseen CLIC energy stages is at the first one above 500\,GeV, assumed to be at 1.4\,TeV in this paper.
In \ZHH production, this energy stage also gives the best sensitivity to \gHHH.
The cross section of WBF double Higgs boson production grows with the collision energy.
Therefore, assuming the same polarisation configuration, the 3\,TeV
stage gives the largest event rate of WBF double
Higgs boson production at CLIC.
In \epem collisions at $\sqrt{s} \gtrsim 1.2$\,TeV, WBF is the dominant double Higgs
boson production mode for unpolarised beams.
Its total cross section at 3\,TeV, including effects of the luminosity spectrum
and initial state radiation, exceeds that of double Higgsstrahlung at
1.4\,TeV by a factor of 6. The single most sensitive measurement of
Higgs boson pair production at CLIC is therefore the double Higgs
boson production through WBF at 3\,TeV.

\begin{figure*}
\includegraphics[width=\textwidth]{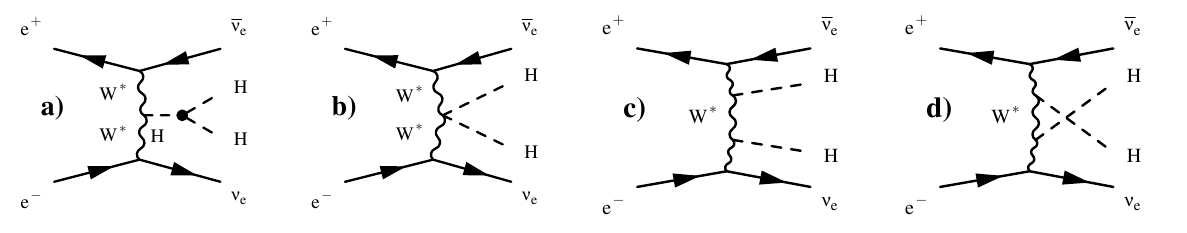}

\caption[Feynman Diagrams for W-boson fusion double Higgs boson production.]{Main Feynman diagrams contributing to double Higgs boson production via W-boson fusion. Diagram a) contains the trilinear Higgs self-coupling,  b) grows with the quartic coupling \gHHWW, while  c) and d) are sensitive to the Higgs coupling to W bosons.}
\label{fig:feynman_WWtoHH}
\end{figure*}

Fig.~\ref{fig:feynman_WWtoHH} shows the main Feynman diagrams
contributing to Higgs pair production via W-boson fusion.
This channel contains the \HHH vertex which depends on the trilinear Higgs self-coupling \gHHH, as well as the \HHWW vertex which depends on the quartic Higgs-gauge coupling \gHHWW.
Deviations from the SM values are defined as:
\[ \kappaHHH := \frac{\gHHH}{\gHHHSM} \mathrm{~and~} \kappaHHWW := \frac{\gHHWW}{\gHHWWSM}. \]

\begin{figure}[htbp]
  \includegraphics[width=0.495\textwidth]{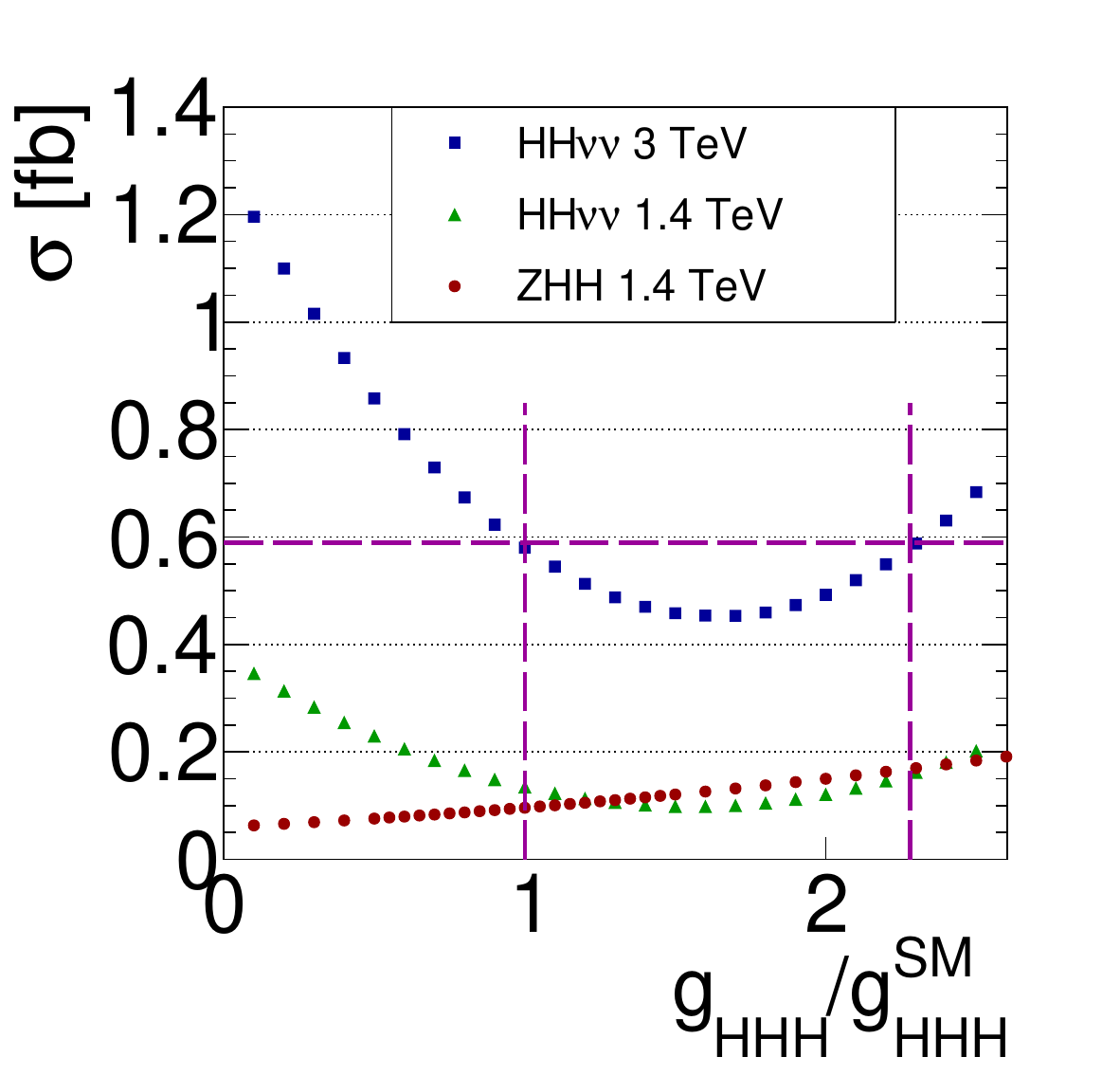}\\

\caption{Cross section dependence of the trilinear Higgs self-coupling
  for the processes \HHvv production at 1.4 and 3\,TeV and \ZHH
  production at 1.4\,TeV for unpolarised beams. Beamstrahlung and
  initial state radiation are included. The SM case is at
  $\frac{\gHHH}{\gHHHSM} =1$. The ambiguity of the cross section value
  in the case of \HHvv production is illustrated for 3\,TeV. No such ambiguity exists in the \ZHH production process.}
\label{fig:xsec_vs_coupling}
\end{figure}

The total cross sections of WBF and Higgsstrahlung double Higgs boson production are sensitive to the value of the trilinear Higgs self-coupling.
Figure \ref{fig:xsec_vs_coupling} shows the parabolic dependence
of the WBF double Higgs boson production cross section on \gHHH at
3\,TeV. The cross section at around $2.3\times \gHHHSM$ is identical to the SM cross section.
Therefore, only measuring the total cross section of this process will not be sufficient to determine \gHHH unambiguously.
This can be resolved by measuring the double Higgsstrahlung cross
section which has an unambiguous dependence on \gHHH as illustrated in
Fig.~\ref{fig:xsec_vs_coupling} for 1.4\,TeV.
Another way to resolve the ambiguity is by using differential
distributions such as the di-Higgs invariant mass~\cite{Contino:2013gna}.
It can also be exploited to distinguish whether a possible deviation from the SM originates from a modification of the \HHH or of the  \HHWW vertex~\cite{Contino:2013gna}.
Differential distributions are therefore used in the following analysis (Sec.~\ref{sec:ResultsFromDifferential}).

This analysis is focused on the two decay channels $\PH\PH\to \bbbb$ (branching fraction 34\,\%) and $\PH\PH\to \bbWW \to \bb \qqqq$ (branching fraction  8.4\,\%).
Both channels benefit from the relatively clean environment in
electron-positron collisions at CLIC, the excellent jet energy resolution of
the assumed CLIC detector concept using particle flow analysis, as
well as from its very good flavour tagging capabilities~\cite{cdrvol2}. This allows
 reconstruction of the kinematic properties of the Higgs boson pair.

The baseline scenario for CLIC sets the collision energy of the second stage to 1.5\,TeV~\cite{Roloff:2645352}.
The earlier choice
of 1.4\,TeV~\cite{CLIC:2016zwp} is used in the present study.
It is expected that prospects for 1.5\,TeV will be very similar
as the cross section only changes by $-7\,\%$ for \ZHH and $+18\,\%$
for \HHvv. 
Results presented here are based on an integrated luminosity of
2.5\,ab$^{-1}$ at a centre-of-mass energy of \SI{1.4}{TeV} and
5\,ab$^{-1}$ at $\sqrt{s}$\,=\,\SI{3}{TeV}.

The CLIC electron beam can be polarised with a polarisation 
of up to  $\pm$80\,\%.
The negative polarisation of $-80\,\%$ leads to an increase of the cross
section for $\epem\to \HHveve$ by a factor of 1.8. The
positive polarisation has the inverse effect of reducing the cross
section to 20\,\%~\cite{Abramowicz:2210491}.
For the process $\epem \to \ZHH$, the cross-section scaling factors are 1.12
(0.88) for the electron beam polarisation of $-80\,\%$ (+80\,\%).
Running a fraction of the integrated
luminosity with positive polarisation is, however, desirable for 
other measurements including two-fermion production~\cite{BSRYR}.
Therefore, a scheme of collecting 80\,\% (20\,\%) of the data with
$-80\,\%$ (+80\,\%) electron beam polarisation is envisaged, which is denoted by ``4:1 polarisation scheme'' in the following. 
A polarisation scaling factor $f_p$ is defined as the ratio of the
total number of events for the assumed polarisation running scheme with respect to the total number of events without beam polarisation for the same total luminosity. 
We apply these scaling factors to obtain the total number of signal and background events for the entire energy stage.
The treatment of the polarisation is detailed in
Sec.~\ref{sec:HHvv-xsec-results}.
A proper optimisation of the selection criteria taking into account
the polarisation dependent kinematics would result in a better signal
selection and hence a higher significance.  The chosen approach is
conservative compared to a proper combination of data sets.

\section{Simulation and reconstruction for signal and background samples}\label{sec:SimulationSamples}
\subsection{Definition of signal and background processes and Monte Carlo generation}\label{sec:sampledef}
The process \epem $\to$ \HHvv with a total cross section of
\SI{0.59}{fb} (\SI{0.149}{fb}) at $\sqrt{s}=$\SI{3}{TeV}
(\SI{1.4}{TeV}) in the decay channels \bbbb and \bbWW defines the
signal.
This includes a contribution from \ZvvHH which
cannot be distinguished from WBF experimentally. It amounts to a fraction of 1.76\,\% of the total $\epem \to \HHvv$ cross section for the unpolarised case at 3\,TeV.
In the baseline polarisation scheme at 1.4 TeV, this contribution is
larger: 13.5\,\% for unpolarised beams, 9.3\,\% and 39\,\% for
negatively and positively polarised electron beams,
respectively. However, these ratios are still small compared to the
statistical uncertainty of the measurement, keeping in mind that the
case of positively polarised beams, which has the largest contribution
of $\PZ(\to \nunubar)\PH\PH$ to the \HHvv final state, only contributes 20\,\% of the luminosity collected at 1.4\,TeV.
The background consists of processes with multiple intermediate
electroweak gauge bosons resulting in multiple jets, single Higgs
boson production in association with electroweak gauge bosons decaying
to hadrons, as well as di-Higgs production with decays to other final states.
In order to avoid overlap, Higgs boson pair production is removed from
the inclusive multi-quark background samples.
Specifically, the background processes which turned out to be non-negligible after the selection are 
$\epem \to \qqqq$ (only relevant at 3 TeV), $\epem \to \qqqq\nunubar$,
$\epem \to \qqqq\Pl \PAGn$, $\epem \to \qqbar \PH \nunubar$,
$\Pe^{\pm} \upgamma \to \PGn \qqqq$, and $\Pe^{\pm} \upgamma \to
\qqbar \PH \PGn$ where \PQq refers to \PQu, \PQd, \PQs, \PQc, and
\PQb quarks, \Pl = $\Pe^\pm$, $\PGm^\pm$, $\PGt^\pm$, and \PGn = \PGne, \PGnGm, \PGnGt,
as well as the respective anti-particles (\PAQq and \PAGn). The processes which do not contain explicitly a Higgs boson in the final state do not include Higgs propagators. 
All SM Higgs boson decays were included otherwise.

Initial state radiation~\cite{Skrzypek:1990qs} and beamstrahlung~\cite{Schulte:1999tx,Poss:1596620} lead to a tail in the
distribution of the effective centre-of-mass energy, which is included
in the simulation.
In addition to \epem collisions, photon-initiated processes are also
considered.
Processes with photons from beamstrahlung in the initial state are
normalised to the corresponding lower luminosity.
``Quasi-real'' photons are modeled using  the Equivalent Photon
Approximation~\cite{vonWeizsacker:1934nji,Williams:1934ad,Budnev:1974de} as implemented in \textsc{Whizard~1.95}~\cite{Kilian:2007gr,Moretti:2001zz}. 

The contributions of the most important background processes are presented in Tables~\ref{tab:HHbbWW_efficiencies_yields_14},~\ref{tab:HHbbWW_efficiencies_yields_3}, ~\ref{tab:HHbbbb_efficiencies_yields_14} and~\ref{tab:HHbbbb_efficiencies_yields_3}.

All samples are generated with \textsc{Whizard~1.95} 
interfaced to
\textsc{Pythia~6.4}~\cite{Sjostrand:2006za} for parton shower and hadronisation as well as Higgs decays. \textsc{Tauola}~\cite{Jadach:1993hs,Jadach:1990mz} is used for $\tau$ lepton decays.

Unpolarised beams are assumed in the simulation samples.
We have studied the effects of polarisation on the kinematics of a
process where a potentially relevant effect is expected: The $\epem \to \PW \PW$ background strongly decreases
with positive electron beam polarisation
as the contribution of the $t$-channel neutrino exchange is
suppressed.
However,
while
the kinematic distributions differ between 100\,\% positive (only
$s$-channel diagrams) and
100\,\% negative ($s$- and $t$-channel diagrams) beam
polarisation,
the contribution from negatively polarised electrons dominates by far in both the 
$P(e^{-})=-80\,\%$ and
$P(e^{-})=+80\,\%$ beam polarisation modes.
Therefore, the \PW boson kinematics are unchanged, such that only the different normalisation between
positive and negative beam polarisation modes has been taken into account in this study.

\subsection{Detector simulation}

The simulation in this analysis uses the CLIC\_ILD detector model~\cite{cdrvol2}.
It is based on the ILD detector concept~\cite{Abe:2010aa,ILD:2020qve} for the
International Linear Collider (ILC) \cite{Behnke:2013lya} adapted to
the experimental conditions at CLIC:
Due to the higher collision energy at CLIC than at the ILC, jets tend
to be more energetic. Therefore, the hadronic calorimeter has more
interaction lengths in CLIC\_ILD than in the ILD concept (7.5 instead
of 5.5\,$\uplambda_{\text{I}}$).
The
magnetic field is slightly higher (4 instead of 3.5\,T).
The inner radius of the vertex detector is 31\,mm in CLIC\_ILD and
16\,mm in ILD.
In addition, the very forward detectors for CLIC\_ILD have been
redesigned as the detector at CLIC is required to cope with more
beam-induced background in particular in the forward region.
The CLIC\_ILD detector has a cylindrical layout. The innermost
subdetector is an ultra-light silicon vertex detector with six layers with a single
point resolution of \SI{3}{\micro\metre}. It is surrounded by a large
tracking system consisting of a large central gaseous Time Projection
Chamber (TPC) surrounded by several silicon strip layers. Highly granular electromagnetic and hadronic calorimeters are located around the tracker. They are optimised for particle flow analysis which aims at reconstructing the final-state particles within a jet using the information from the tracking detectors combined with that from the calorimeters. 
The outermost part of the detector consists of an iron return yoke,
which is instrumented with muon chambers. The forward region is
equipped with a system of two electromagnetic calorimeters, the
BeamCal and LumiCal. They are specifically designed for the luminosity measurement and the identification of electromagnetic clusters from forward electrons or photons.

At CLIC, the bunch crossings are separated by \SI{0.5}{ns}.
At the 3\,TeV stage, there are
on average 3.2 \gghad  interactions per bunch crossing~\cite{cdrvol2}.
In order to suppress the beam-induced background collected over the duration of a bunch train,
the hit time resolution in the calorimeters is \SI{1}{ns} while the
TPC integrates over the entire bunch train.
The elements of the silicon envelope of the TPC and the vertex
detector have a time resolution of $10/\sqrt{12}$\,ns.

Recently, a new detector model, CLICdet, has been optimised and validated for CLIC~\cite{Arominski:2018uuz}. The performance of this analysis is expected to be similar if the CLICdet model had been used.

The detector simulation of the generated event samples is performed
with \textsc{Geant4}~\cite{Agostinelli2003,Allison2006} and the
detector description toolkit
\textsc{Mokka}~\cite{MoradeFreitas:2002kj}.
Hits from 60 bunch crossings of beam-induced \gghad background are overlaid to each event.
This is done for all subdetectors. For most of them, this is more than the reconstruction window and hit resolution requires. For the TPC, this is a compromise between realism and computing capacities~\cite{Schade:1443537,cdrvol2}.

\subsection{Reconstruction}

The reconstruction algorithms run in the \textsc{Marlin} framework~\cite{MarlinLCCD} which is a part of iLCSoft~\cite{ILCSoft}.
This includes track reconstruction with the ILD track reconstruction software~\cite{1742-6596-513-2-022011} and particle flow analysis based on tracks and calorimeter deposits with the \textsc{PandoraPFA} program~\cite{THOMSON200925,Marshall:2012ry,Marshall:2015rfa} resulting in Particle Flow Objects (PFOs).
Cuts on the timing of the PFOs are applied to suppress beam-induced
backgrounds from other bunch crossings.
Muon and electron candidates are identified using calorimeter and
tracking information.
They are required to be isolated 
by applying quality criteria on their impact parameters and by restricting the  energy in the surrounding cone in dependence on the track energy.
As the forward calorimeters were not used in the reconstruction, the geometrical acceptance and the efficiency of the forward
calorimeters BeamCal and LumiCal from dedicated full simulation~\cite{Sailer:2017onh} are used to
simulate the veto of forward electrons occurring in background processes
in the polar angle region between 10 and 110\,mrad.

Jets are reconstructed using the
\textsc{FastJet}~\cite{Cacciari:2011ma} package via the
\textsc{MarlinFastJet} interface.
Both the VLC algorithm~\cite{Boronat:2014hva,Boronat:2016tgd}\footnote{Slightly
  differing from the definition given in~\cite{Boronat:2016tgd}, the
  beam distance is determined as $d_{i\text{B}} =
  E_i^{2\beta}(p_{\text{T},i}/E_i) ^{2\gamma}$ instead of
  $d_{i\text{B}} = E_i^{2\beta} \sin^{2\gamma}\theta_{i\text{B}}$.}
and the longitudinally invariant $k_{t}$
algorithm~\cite{Catani:1991hj} are used in the analysis.
The parameter settings for the jet reconstruction in the individual
channels are specified in Sec.~\ref{sec:eventselection}.
Vertex reconstruction and heavy-flavour tagging is performed using the
Linear Collider Flavour Identification (\textsc{LcfiPlus}) program~\cite{Suehara:2015ura}.
Hadronic tau decays are identified using the TauFinder package~\cite{Muennich:1443551}.

The jets studied in this paper are predominantly $b$-jets with an energy around 100\,GeV which are rather forward in the detector.
The pure relative jet energy resolution achievable with the CLIC\_ILD
detector is between 3 and 5\,\% for light-flavour jets~\cite{cdrvol2,Marshall:2012ry}.
However, for forward $b$-jets such as those in this analysis, the
resolution is degraded for several reasons:
A part of the jet energy is missing due to neutrinos from heavy
flavour decays and due to forward particles outside of the detector
acceptance. In addition, the beam-induced background is higher in the
forward region.

The momentum resolution for forward tracks with a transverse momentum around 100\,GeV  is estimated to be around $\sigma(\Delta p_{\text{T}}/p_{\text{T}}^2) =
9\times 10^{-4}$\,GeV$^{-1}$~\cite{cdrvol2} 
and the impact parameter resolution
is $\sigma_{d_0} \approx 1.5\,\upmu$m for central tracks and
around $\sigma_{d_0} \approx 3\,\upmu$m in the forward region~\cite{cdrvol2}. 
The $b$-tagging performance is expected to provide a
mis-identification rate of around 0.1\,\% for light flavor jets with a
$b$-tagging efficiency of 55\,\%~\cite[Fig.\,6]{Abramowicz:2018rjq}
for jets with a similar polar angle distribution as the signal.
\section{Event selection}
\label{sec:eventselection}
\subsection{Common preselection and definition of orthogonal samples}\label{sec:presel-orthogonality}
To select events  originating from double Higgs production
in the \bbbb~and \bbWW$\to \bb \qqqq$~decay channels, all events
containing isolated leptons (electrons, muons or hadronic $\tau$ leptons) are rejected.
For this, electron and muon candidates compatible with prompt
production with an absolute impact
parameter below 0.04\,mm (0.06\,mm) for electrons
(muons) are used. Furthermore, the fraction of energy deposited in the
electromagnetic calorimeter $R_{\text{cal}}$ is required to be
$R_{\text{cal}} > 0.9$ for electrons and $0.05 < R_{\text{cal}} <
0.25$ for muons. A minimum track energy of 15\,GeV is required, and an energy-de\-pendent
cone-based isolation criterion is imposed, allowing a typical maximum energy in
the cone of, \eg, 23\,GeV for 100\,GeV tracks.
For the identification of hadronically decaying $\tau$ leptons, parameters are chosen to optimise the performance for this analysis. In particular, a maximum
energy of 3\,GeV in the cone between 0.03 and 0.33\,rad around the
seed particle is required.

\begin{figure}
   \includegraphics[width=0.45\textwidth]{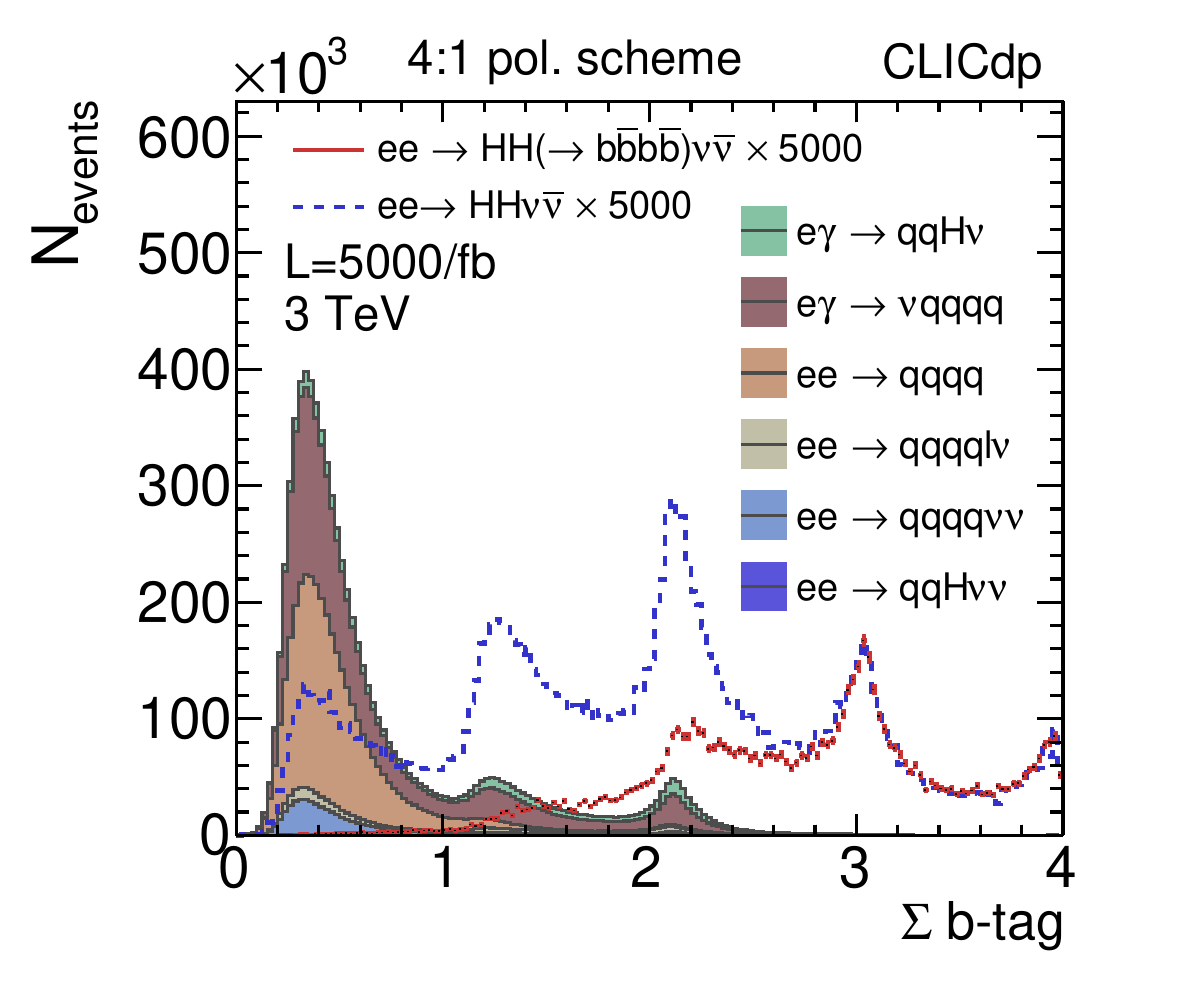}
\caption[sumbtag]{Distribution of the sum of the $b$-tag values for
  the inclusive \HHvv and the \HHvv $\to$ $\bbbb\nunubar$ channel,
  both scaled by a factor $5000$ for better visibility, and for the background processes. No selection is applied.}
\label{fig:sumbtag}
\end{figure}

In order to define orthogonal samples to be used for the \bbbb~and
\bbWW~channels, the events are clustered into four jets using the
\kT~algorithm with a jet size parameter of $R=0.7$.
A flavour tagging algorithm is applied on these jets using the
\textsc{LcfiPlus} package.
It first identifies the primary vertices, followed by the secondary vertices indicating $b$ and $c$ hadron decays. Then, the secondary vertices are assigned to jets.  In the next step, the jet clustering is refined by using as seeds only those tracks and leptons originating from secondary vertices. Finally, values for $b$ tags, $c$ tags, and light-flavour quark tags are assigned to each jet. This classification is based on a multivariate discriminant trained on $\epem \to \PZ \nunubar$ events, which have a similar event topology to the signal events.
In the training, events with Z bosons subsequently decaying to \bb are treated as signal, while those decaying to either \cc or \qqbar (with q = \PQu, \PQd, \PQs) are considered background. 
The $b$-tagging performance relies on the ability to identify secondary vertices and tracks which do not originate from the primary interaction point. 
This depends in particular on the single point resolution of the vertex detector.
In the CLIC\_ILD model, this is assumed to be $\approx$ 3\,$\upmu$m. Flavor tagging performances reached with the CLIC\_ILD detector at 1.4\,TeV are illustrated in~\cite[Fig. 6]{Abramowicz:2018rjq}.
The  $\sum_{4}{b\text{-tag}} $ distribution at \SI{3}{TeV} is shown in
Fig.~\ref{fig:sumbtag}.
The \bbbb final state tends to be in the region between 2 and 4 of the $\Sigma_4$$b$-tag distribution, which is much higher than the backgrounds. Contributions from other Higgs decays tend to values between 0 and 2.5.
This shows that this criterion can be used
to remove background contributions, and a large contribution of other $\PH\PH$ decay channels is also removed.
The sample is then split into mutually exclusive samples with \bbbb~and \bbWW~candidates in the following way:
Events are chosen as \bbWW~candidates if the sum of the $b$-tag values $\sum_{4}{b\text{-tag}} $ of the jets is smaller than 1.5 (2.3) at \SI{1.4}{TeV} (\SI{3}{TeV}).
Otherwise, the events are considered as \bbbb~candidates.
Further selection criteria are applied separately for the two channels.

\subsection{Double Higgs production in the decay to \bbWW}

In the \bbWW decay channel, the fully leptonic and semi-leptonic final
states are dominated by background processes with leptons and missing transverse momentum~\cite{Xu:2293892}. Therefore, only the fully hadronic final state is considered here.
The analysis is optimised separately for 1.4 and 3\,TeV.

After the initial classification, 
the  candidate events for \bbWW $\to$ $ \bb\qqqq$ are re-clustered into six
jets using the longitudinally invariant $k_{t}$ algorithm with a radius parameter of $R=0.7$.
The six jets are grouped by minimising
\begin{equation}
  \chi^{2} = \frac{(m_{ij} - m_{\PH})^{2}}{\sigma^2_{\PH\to \bb}} + \frac{(m_{klmn} - m_{\PH})^{2}}{\sigma^2_{\PH\to \PW\PW^{*}}} + \frac{(m_{kl} - m_{\PW})^{2}}{\sigma^2_{\PW}} ,
  \label{eq:bbww:chisqu}
\end{equation}
where $i, j, k, l, m, n$ are the indices denoting the six jets.
The invariant mass resolutions are obtained from fitting a
Gaussian-like function with asymmetrical width parameters to the
respective peaks in the invariant mass spectra of the decay products,
obtaining   
$\sigma_{\PH\to \bb}$ = \hfill\\15.0\,GeV (8.4\,GeV), $\sigma_{\PH\to
  \PW\PW^{*}}$ = 36.6\,GeV (7.4\,GeV), and $\sigma_{\PW}$ = 13.1\,GeV
(9.5\,GeV) at 3\,TeV for the width below (above) the maximum, and similar
values at 1.4\,TeV~\cite{Xu:2293892}.
To suppress background processes without $b$-quarks while minimising signal loss, the highest $b$-tag value among the six jets has to be at least 0.7 at 3\,TeV. At 1.4\,TeV, the second highest $b$-tag value is required to be above 0.2.
As the contribution of $s$-channel processes such as $\epem \to \qqqq$ compared to \PW-boson fusion processes is larger at 1.4\,TeV, in addition the total transverse momentum of the Higgs boson pair is required to be larger than 30\,GeV, which enhances the fraction of processes with neutrinos in the final state~\cite{Xu:2293892}.

The signal selection is performed using Boosted Decision Trees (BDTs) trained
on the following input variables~\cite{Xu:2293892}:
 Invariant masses and angular distributions of the \bb system, of the \WW system, of the jets associated with the W decay, and of the
 \bbWW system, as well as the energy of the jets originating from the
 W boson are provided to the training.
In addition, the transverse momenta of the two reconstructed Higgs
bosons and of the di-Higgs system, angular variables between the  rest-frames of the \bb,
\WW and HH systems and of the jets associated with the W decays as
well as the sphericity, the merging scales of exlusive jet
clustering, and $b$- and $c$-tag values are used as input to the BDT
training as well.
A cut on the BDT response is applied to maximise the precision of the
cross section measurement.
The resulting event yields in the signal region for the HH$\to \bbWW$
signal and the main background processes are listed in Table~\ref{tab:HHbbWW_efficiencies_yields_14} for 1.4\,TeV and in  Table~\ref{tab:HHbbWW_efficiencies_yields_3} for 3\,TeV.
Although the signal selection is optimised for the decay channel \bbWW, there are significant contributions from other Higgs decay channels as well.

Comparing the efficiencies between the two collision energies, the
Higgs bosons in the
\HHvv signal become more forward at 3\,TeV.
This is one of the reasons why the processes with $\Pe^{\pm}\PGg$
initial state are kinematically more similar to the signal at 3\,TeV, making it
more difficult to suppress them.

\begin{table}[ht]
  \caption{Cross sections, $\sigma$, selection
    efficiencies, $\epsilon_{\text{BDT}}$, and expected
    number of events in the HH$\to
    \bbWW$ signal region, $N_{\text{BDT}}$, at $\sqrt{s}
    =$\SI{1.4}{TeV} for $\mathcal{L}=2.5$\,ab$^{-1}$. The cross
    sections are for unpolarised beams; the number of events assumes the 4:1 polarisation scheme~\cite{Abramowicz:2210491}.}

  \begin{tabular}[ht]{lcrrccc}

\toprule
    Process & $\sigma$/fb & $\epsilon_{\mathrm{BDT}}$
    & $N_{\mathrm{BDT}}$ \\\midrule
    \HHvv; HH $\to \bbWW$;& \multirow{2}{*}{0.018}& \multirow{2}{*}{4.9\,\%} & \multirow{2}{*}{3}  \\
~~~    $\PW\PWst \to \qqqq$ & \\
    
    \HHvv; HH $\to \bbbb$  & 0.047 &0.075\,\% & 0.1\\
    \HHvv; HH $\to $ other & 0.085 &0.34\,\% &1.1\\
\midrule
    $\epem \to \qqqq\nunubar$  &23 & 0.00034\,\% & 0.3  \\
    $\epem \to \qqqq\Pl \PAGn$  & 110 & 0.001\,\%& 4 \\
    $\epem \to \qqbar  \PH \nunubar $& 1.5 & 0.035\,\%& 1.9    \\
    $\Pe^{\pm} \upgamma \to \nu \qqqq$   & 154 & 0.001\,\% & 6     \\
    $\Pe^{\pm} \upgamma \to  \qqbar \PH \PGn $ & 30 &0.005\,\%& 6\\

\bottomrule

  \end{tabular}

  \label{tab:HHbbWW_efficiencies_yields_14}
\end{table}

         \begin{table}[ht]
  \caption{
Cross sections, $\sigma$, selection
    efficiencies, $\epsilon_{\text{BDT}}$, and expected
    number of events in the HH$\to
    \bbWW$ signal region, $N_{\text{BDT}}$, at $\sqrt{s}
    =$\SI{3}{TeV} for $\mathcal{L}=5$\,ab$^{-1}$. The cross
    sections are for unpolarised beams; the number of events assumes the 4:1 polarisation scheme~\cite{Abramowicz:2210491}.}
  \begin{tabular}[ht]{lcrrccc}

\toprule
    Process & $\sigma$/fb & $\epsilon_{\mathrm{BDT}}$
    & $N_{\mathrm{BDT}}$ \\\midrule

    \HHvv; HH $\to \bbWW$;  & \multirow{2}{*}{0.07}                    &   \multirow{2}{*}{ 7.4\,\%} &\multirow{2}{*}{ 38 } \\
~~~    $\PW\PWst \to \qqqq$ & \\
    \HHvv; HH $\to \bbbb$  & 0.19                    &   0.28\,\% & 4   \\
    \HHvv; HH $\to $ other &0.34                     &    0.72\,\% & 18  \\
\midrule
    $\epem \to \qqqq$                 &547                      &    0.00014\,\% & 6  \\
    $\epem \to \qqqq\nunubar$    & 72                      &    0.0045\,\% & 24   \\
    $\epem \to \qqqq\Pl \PAGn$   &107                      &   0.0037\,\% & 29  \\
    $\epem \to \mathrm{q} \bar{\mathrm{q}} \mathrm{ H} \nunubar$  &4.8  &  0.19\,\% & 68  \\
    $\Pe^{\pm} \upgamma \to \nu \qqqq$         &523                      &     0.006\,\% & 232  \\
    $\Pe^{\pm} \upgamma \to \qqbar \PH \PGn $ &116                     &    0.054\,\% & 463  \\ 

\bottomrule

  \end{tabular}
    
  \label{tab:HHbbWW_efficiencies_yields_3}
\end{table}

\subsection{Double Higgs production in the decay to \bbbb}\label{sec:bbbb}

Candidate events for the final state \bbbb at 3\,TeV are pre-selected
according to the orthogonality selection
(Sec. \ref{sec:presel-orthogonality}).
The events are re-clustered
with the VLC algorithm which provides an improvement in the di-jet
mass resolution as shown in~\cite{Boronat:2016tgd}.
The VLC algorithm is applied 
in exclusive mode requiring $N=4$ jets and
using a radius parameter $R=1.1$ and the parameters $\beta = \gamma = 1$.
To enhance the signal fraction at $\sqrt{s} = $\SI{1.4}{TeV}, if
$\sum_{4}{b\mathrm{-tag}} < 2.3$, events are required to have a sum of
the jet energy of $\sum{E(\text{jet})}>$ \SI{150}{GeV} and the second
highest jet transverse momentum must be $p_{T}(\text{jet}_{2}) > $\SI{25}{GeV}.

Since both Higgs bosons are expected to be on-shell, the four jets are then grouped as two Higgs candidates by minimising the absolute difference between the resulting di-jet masses $|m_{ij} - m_{kl}|$.
BDTs are trained based on the pre-selected events in order to optimise the signal selection efficiency and purity.

The following observables were chosen for the multivariate analyses: the sum of all $b$-tag weights,
the ratio between the sum of all $c$-tag weights and the sum of all $b$-tag weights, 
the invariant mass of each jet pair, 
the cosine of the angle between the two paired jets for each jet pair evaluated in the centre-of-mass system, 
the total invariant mass of the system, 
the missing transverse momentum computed as the opposite of the vectorial sum of the momenta of all jets, 
the number of photons with energy larger than \SI{25}{GeV}, and the
maximum absolute pseudorapidity among the four jets.
These ovservables are sensitive to various properties distinguishing
the signal from background processes such as the presence of heavy
flavour jets and neutrinos, invariant mass and angular distributions
of Higgs boson decay products, and to differences of the \PW-boson
fusion to the $s$-channel topology.
The analyses are optimised separately for 1.4 and 3\,TeV.

\begin{table}[ht]
  \caption{
Cross sections, $\sigma$, selection
    efficiencies, $\epsilon_{\text{BDT}}$, and expected
    number of events in the HH$\to
    \bbbb$ signal region, $N_{\text{BDT}}$, at $\sqrt{s}
    =$\SI{1.4}{TeV} for $\mathcal{L}=2.5$\,ab$^{-1}$. The cross
    sections are for unpolarised beams; the numbers of events assume the 4:1 polarisation scheme~\cite{Abramowicz:2210491}.}

  \label{tab:HHbbbb_efficiencies_yields_14}
  \begin{tabular}[ht]{lcccccc}
 
    \toprule
  Process & $\sigma$/fb &  $\epsilon_{\mathrm{BDT}}$ & $N_{\mathrm{BDT}}$  \\\midrule
     
    $\epem \to$ \HHvv                       & 0.149 &  7\,\%& 40 \\
    \midrule                                                      
     ~~~~ only HH$\to \bbbb$                                       & 0.047 &  23\,\% & 39\\  
     ~~~~ only HH$\to$ other                                       & 0.102 &  0.22\,\%& 0.8\\ \midrule
    $\epem \to \qqqq\nunubar$                           & 23 & 0.02\,\%& 20 \\
    $\epem \to \qqqq\Pl \PAGn $                          & 110 &0.005\,\% & 19 \\
    $\epem \to \qqbar \PH \nunubar $ & 1.5 &0.8\,\%  &43  \\
    $\Pe^{\pm} \upgamma \to \PGn \qqqq$
          & 154 &0.0013\,\% &7 \\ 
     $\Pe^{\pm} \upgamma \to \qqbar \PH \PGn  $                       & 30 & 0.003\,\%&3 \\
 \bottomrule
    
  \end{tabular}

\end{table}

\begin{table*}[ht]
  \caption{Cross sections, $\sigma$, selection
    efficiencies, $\epsilon_{\text{looseBDT}}$ ($\epsilon_{\text{tightBDT}}$), and expected
    number of events in the loose (tight) BDT selection region of the HH$\to
    \bbbb$ analysis, $N_{\text{looseBDT}}$ ($N_{\text{tightBDT}}$), at $\sqrt{s}
    =$\SI{3}{TeV} for $\mathcal{L}=5$\,ab$^{-1}$. The cross
    sections are for unpolarised beams; the numbers of events assume the 4:1 polarisation scheme.}
\centering
  \begin{tabular*}{\linewidth}{@{\extracolsep{\fill}}*1l@{}@{\extracolsep{\fill}}*6c@{}}

    \toprule
  Process & $\sigma$/fb &  $\epsilon_{\mathrm{looseBDT}}$ & $N_{\mathrm{looseBDT}}$  & $\epsilon_{\mathrm{tightBDT}}$ & $N_{\mathrm{tightBDT}}$\\\midrule

      $\epem \to$ \HHvv                       &
                         0.59 &17.6\,\% &  766&  8.43\,\%& 367\\
    \midrule
      ~~~~ only HH$\to \bbbb$                               & 0.19 &
                                                                     53.4\,\%
                                                          &734 & 26.3\,\%
                                                                                                                      & 361\\
     ~~~~ only HH$\to$ other                               & 0.40 &
                                                                    1.1\,\%
                                                          &32&0.2\,\% & 6    & \\\midrule
     $\epem \to \qqqq$                                        & 547  &0.0065\,\% &259 &0.00033\,\%&  13\\
     $\epem \to \qqqq\nunubar$                           & 72   &0.17\,\% &876 &0.017\,\% &  90\\
     $\epem \to \qqqq\Pl \PAGn$                          & 107  & 0.053\,\% &421 &0.0029\,\% &  23\\
     $\epem \to \qqbar  \PH \nunubar$ & 4.7  &3.8\,\% & 1171& 0.56\,\%& 174\\
     $\Pe^{\pm} \upgamma \to \PGn \qqqq$                                & 523  &0.023\,\% &821 &0.0014\,\% &  52\\
     $\Pe^{\pm} \upgamma \to \qqbar \PH \PGn  $                       & 116  & 0.12\,\% &979 & 0.0026\,\%& 21\\
 \bottomrule
  
  \end{tabular*}
    
  \label{tab:HHbbbb_efficiencies_yields_3}
\end{table*}

For the cross section measurement, the cut on the BDT response is optimised for the signal significance.
The resulting expected event yields for the 1.4\,TeV analysis are listed in Table~\ref{tab:HHbbbb_efficiencies_yields_14}.
At 3\,TeV, two selections are defined: the ``tight BDT'' region with a
BDT cut of BDT response $\,>\,0.12$, which is optimised for signal significance,
and the ``loose BDT'' region with a cut of BDT$\,>\,0.05$, which is
optimised for the extraction of the Higgs self-coupling.
The expected event yields for the two selection variants at 3\,TeV for a
luminosity of $\mathcal{L}=5$\,ab$^{-1}$  are
listed in 
Table~\ref{tab:HHbbbb_efficiencies_yields_3}.
Both selection regions contain also a significant contribution from decays other than \bbbb.
As the processes with $\Pe^{\pm}\PGg$ initial state do not produce
more than 2 $b$-jets, they are strongly suppressed by criteria based
on the $b$-tag weights. This makes the fraction of these backgrounds passing the event selection less energy-dependent than in the \bbWW analysis.
This is also reflected in the fact that the BDT selection is more efficient for
the signal in the \bbbb analysis at both energies.

\begin{figure}
  (a)\\
  \includegraphics[width=0.49\textwidth]{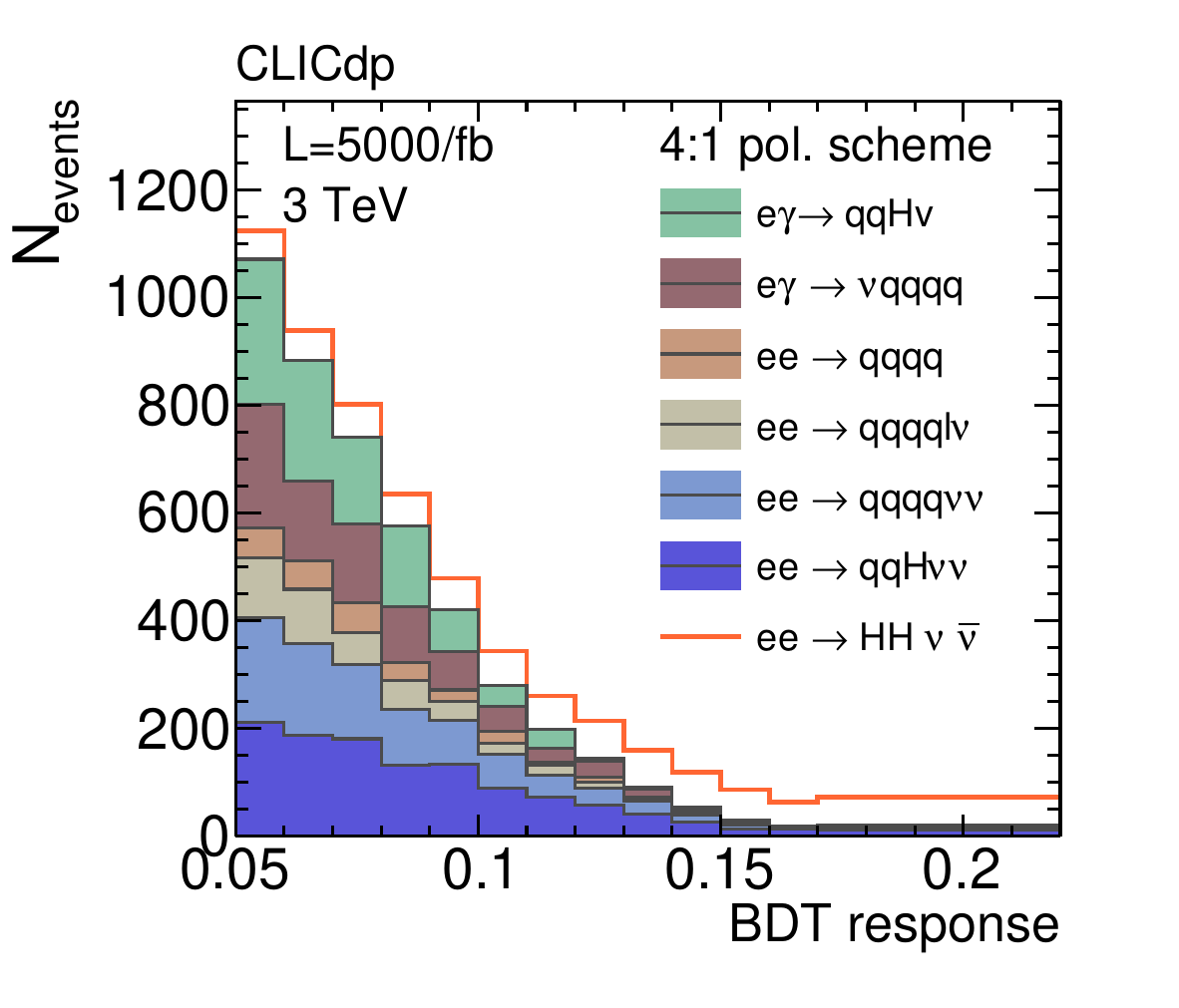}\\
  (b)\\
\includegraphics[width=0.49\textwidth]{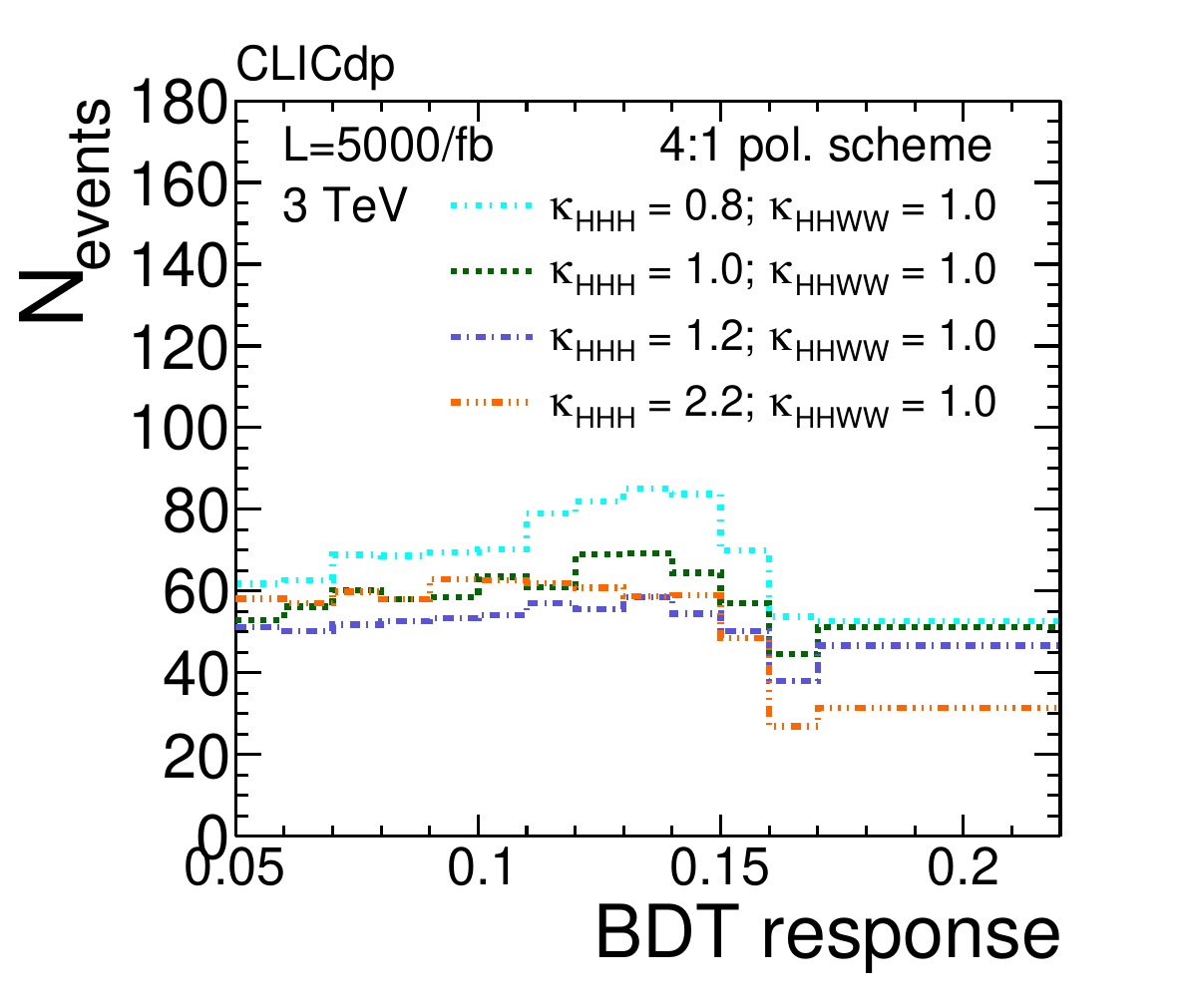}
\caption[BDT]{BDT response distribution of (a) all SM contributions
  stacked and (b) a selection of signal samples with modified \gHHH in
  the loose BDT selection at 3\,TeV CLIC. The Higgs-gauge boson coupling
  \gHHWW is kept at its SM value.}
\label{fig:BDT_withsignal}
\end{figure}

\begin{figure}[ht]\centering
  \includegraphics[width=0.49\textwidth]{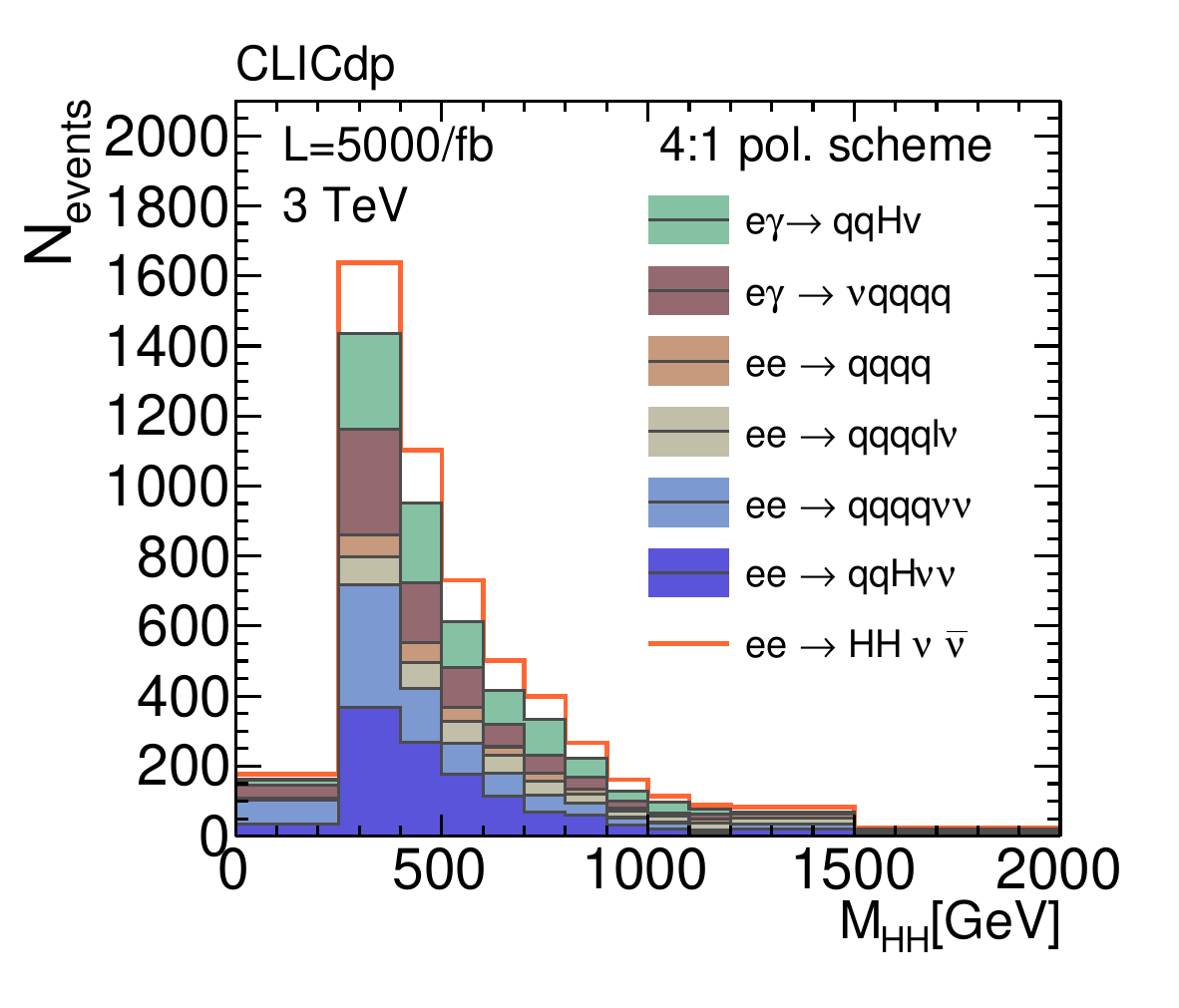}
\caption[BDT]{Invariant mass of the Higgs boson pair for the  SM contributions in the loose BDT selection at 3\,TeV CLIC.}
\label{fig:SumMj_looseBDT_withsignal}
\end{figure}

\begin{figure}(a)\\
  \includegraphics[width=0.49\textwidth]{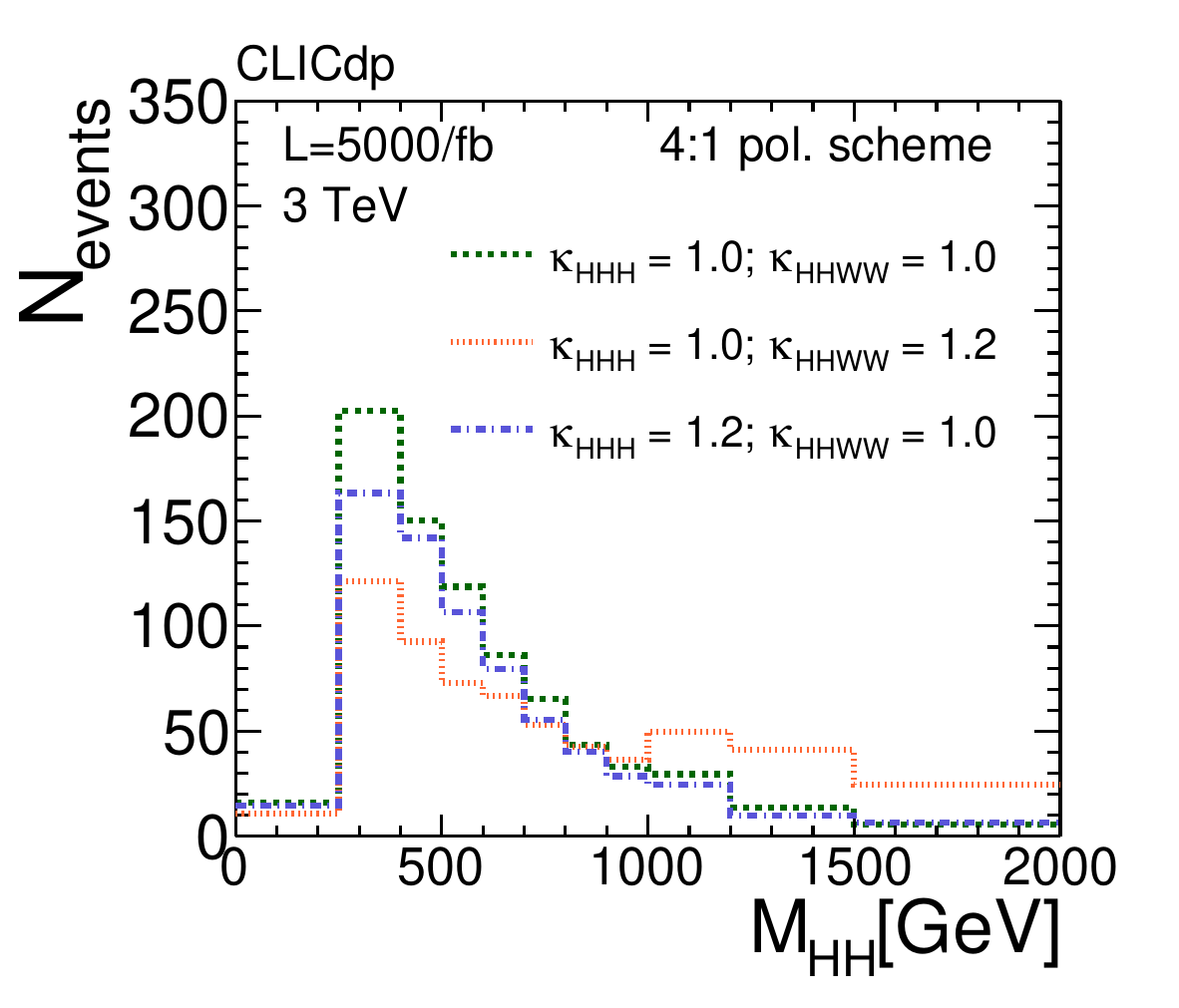}\\
(b)\\ \includegraphics[width=0.49\textwidth]{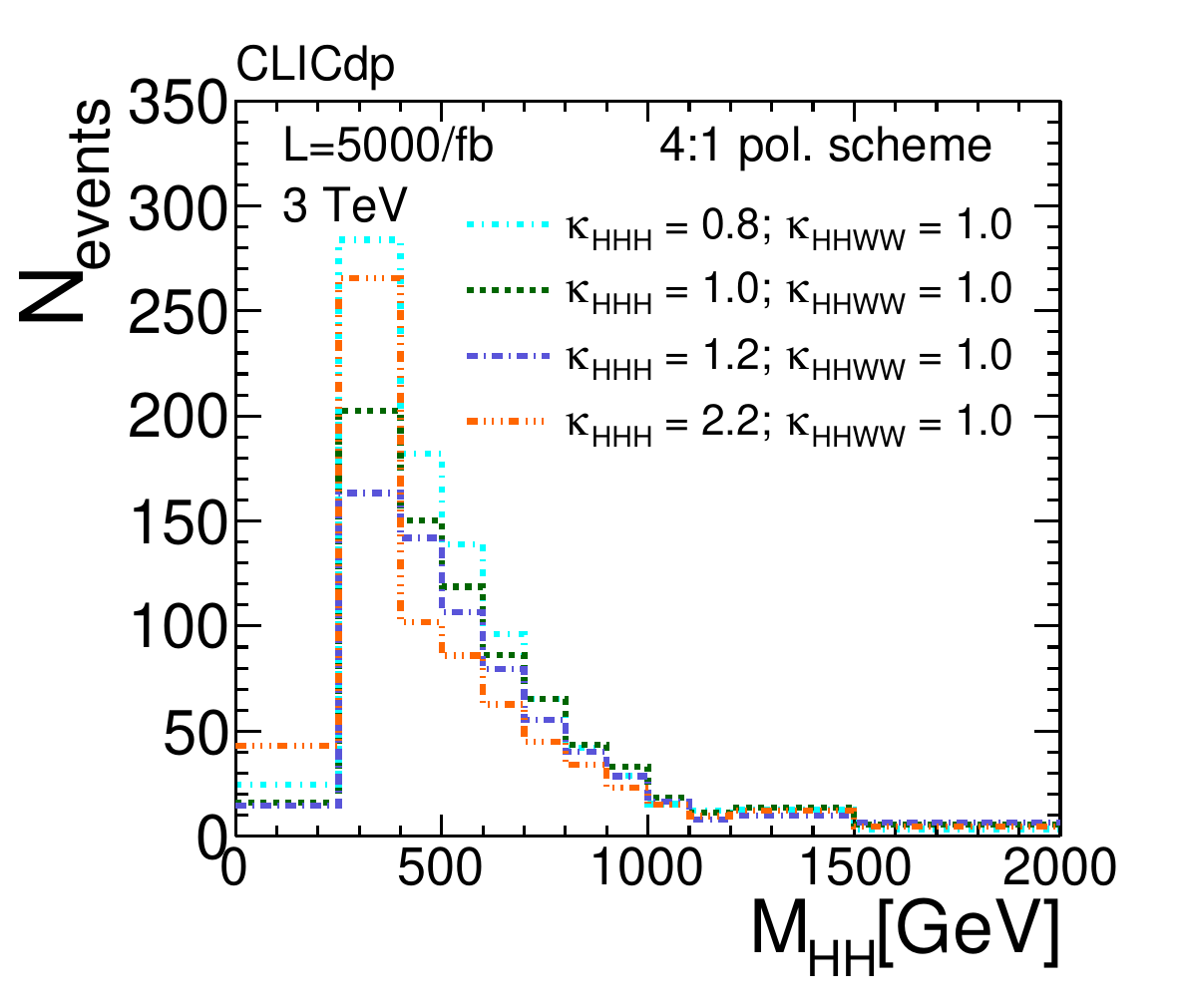}
\caption[]{Invariant mass of the Higgs pairs for the signal contributions with different values of \kappaHHH and \kappaHHWW in the loose BDT region. (a) Comparing samples with one of the couplings fixed to the SM value. (b) Comparing samples with $\kappaHHH<1$ and $\kappaHHH>1$. The sample with  $\kappaHHH=2.2$ has roughly the same total cross section as the SM case.}
\label{fig:SumMj_SignalsOnly_7181_7183_8201}
\end{figure}

While the cross section is measured in the tight BDT region, the expected precision on \gHHH and \gHHWW is evaluated  based on differential distributions in the loose BDT region to allow for a larger event sample.
Fig.~\ref{fig:BDT_withsignal} shows the distribution of the BDT
response in the loose BDT region.
From Fig.~\ref{fig:BDT_withsignal}\,(a), it can be seen that the SM \HHvv
signal is dominant compared to backgrounds at higher BDT score values.
Selected samples with modified \gHHH are compared in 
Fig.~\ref{fig:BDT_withsignal}\,(b), which shows a small overall
sensitivity of the BDT score to the Higgs self-coupling.
The main influence on the area of the distributions is the total cross
section: The selection efficiencies vary only between 17 and 18\,\% among the event
samples with the given coupling values, while the total
cross sections vary between 0.471 and 0.68\,fb
The distribution of the invariant mass of the double Higgs boson system for the SM contributions in the loose BDT region is presented in Fig.~\ref{fig:SumMj_looseBDT_withsignal}.
Fig.~\ref{fig:SumMj_SignalsOnly_7181_7183_8201}\,(a) and \ref{fig:SumMj_SignalsOnly_7181_7183_8201}\,(b) show the invariant di-Higgs mass distributions for  selected values of \gHHH and \gHHWW.
The di-Higgs invariant mass distributions between points with
similar, but opposite, variation of the \gHHH coupling differ
especially in the lower invariant mass region as illustrated in
Fig.~\ref{fig:SumMj_SignalsOnly_7181_7183_8201}\,(b), comparing the
distributions between $\kappaHHH = 0.8, 1.2$ and $2.2$ to the SM.
As shown in Fig.~\ref{fig:SumMj_SignalsOnly_7181_7183_8201}, the
\gHHWW coupling impacts also the higher invariant mass region, which
allows it to be distinguished from modifications in the \gHHH coupling.

\section{Cross section measurement}\label{sec:ResultsFromXSec}

\subsection{Precision of the cross section measurement for
  \HHvv production at 1.4 and 3\,TeV}\label{sec:HHvv-xsec-results}

The cross-section measurement is based on the baseline luminosity and polarisation scheme resulting in the
event yields for the WBF Higgs pair production signal and the backgrounds listed in Tables~\ref{tab:HHbbWW_efficiencies_yields_14} and~\ref{tab:HHbbWW_efficiencies_yields_3} for the \bbWW analysis and in Tables~\ref{tab:HHbbbb_efficiencies_yields_14} and~\ref{tab:HHbbbb_efficiencies_yields_3} for the \bbbb analysis.
From this, the precision of the cross-section measurement assuming the
SM value can be determined according to $\frac{\Delta \sigma}{\sigma}
= \frac{\sqrt{S+B}}{S}$,
where $S$ ($B$) is the number of signal (background) events passing the selection.
In the \bbbb (\bbWW) analysis channel, the contribution of $\PH\PH$
decaying to other final states than \bbbb (\bbWW) yet passing the signal selection is counted towards the number of signal events.

The \ZvvHH contribution to the \HHvv final state
        exhibits a dependence on \gHHH, though a different one than
        the WBF component. In the \HHvv analysis at 3\,TeV, the
        modification of the \HHvv cross section due to the variation
        of \gHHH is treated as independent of a possible
        change in the \ZvvHH contribution due to analysis selection
        criteria. It has been checked that the impact of a different
        efficiency for the \ZvvHH component is small. In a future
        study, the components could be
        separated in the signal region based on kinematic information, using their individual
        dependencies on \gHHH.

  The energy stage at $\sqrt{s}=\SI{1.4}{\TeV}$ with an integrated
  luminosity of
  $\mathcal{L}$\,=\,2.5\,ab$^{-1}$  and the 4:1
  polarisation scheme provides evidence for the 
  $\epem\to \HHvv$ process with a measurement significance of 3.5\,$\sigma$ corresponding to a cross-section precision of 28\,\%.
At the 3\,TeV stage alone, the observation of $\epem\to \HHvv$ production  is reached after 700\,fb$^{-1}$ of data taking.
 Based on the 3\,TeV stage and both decay channels, the precision of
 the \HHvv cross-section measurement is 7.3\,\%.
The 3\,TeV stage clearly dominates the cross-section measurement for WBF double Higgs production.
With the \bbbb channel at 3\,TeV alone, the precision is 7.4\,\%.
This demonstrates that  the contribution from the \bbWW analysis is
very small. In the following, we therefore consider only the \bbbb analysis.
The uncertainties on the cross section measurement are summarised in
Table~\ref{tab:sigma-unc}.

\begin{table}[ht]
  \caption[Measurement uncertainties of the \HHvv cross section at
  different  stages of CLIC]{Measurement uncertainties for the cross section of
    $\epem \to \HHvv$ at the different stages of CLIC with different
    collision energy $\sqrt{s}$ and integrated luminosity $\mathcal{L}$ including
    different decay channels and assuming the 4:1 polarisation scheme.}
  \label{tab:sigma-unc}
  \begin{tabular}[ht]{lllr}
\toprule $\sqrt{s}$ &$\mathcal{L}$ &decay channel(s) &$\frac{\Delta[\sigma(\HHvv)]}{\sigma(\HHvv)} $\\\midrule
1.4\,TeV &2.5\,ab$^{-1}$ & \bbbb \& \bbWW &28\,\% \\
3\,TeV &5\,ab$^{-1}$  &  \bbbb &  7.4\,\% \\
    3\,TeV& 5\,ab$^{-1}$ & \bbbb \& \bbWW &  7.3\,\%\\
    \bottomrule

  \end{tabular}

\end{table}

As described in Sec.~\ref{sec:strategy}, the $\epem\to \HHvv$
cross section is dependent on the beam polarisation.
In the nominal 4:1 polarisation scheme, the number of $\epem\to \HHvv$
events is scaled by
a factor of $f_p=1.43$ $(1.48)$ at 1.4\,TeV (3\,TeV).
For the \Zhh process at 1.4\,TeV, the polarisation factor is $f_p= 1.072$.
The background composition depends on the electron beam polarisation
modes as well.
As described in Sec.~\ref{sec:sampledef}, the background kinematics
have been found to be mostly independent of the polarisation.
Therefore, unpolarised beams are used for the simulation and the
polarisation is only taken into account in the cross section of the
background processes.
Some of the backgrounds scale by the same polarisation factor as the signal, others are
influenced less by the polarisation. We scale
all backgrounds by the same factor $f_p=1.48$.
This constitutes an upper limit for the background in the negative polarisation run and a lower limit in the positive polarisation run.
Since overall  more luminosity is collected with negative beam polarisation, this is a conservative approach.
Table~\ref{tab:xsec-unc-pol-lumi} shows the dependence of the \bbbb cross-section
measurement uncertainty on the  polarisation.

\begin{table}[ht]
  \caption{Dependence of the cross-section measurement for
    \HHvv at 3\,TeV on the distribution of the luminosity between the two beam polarisation states of the electron beam.
The same polarisation factor is
  assumed for signal and background as explained in the text.}
  \label{tab:xsec-unc-pol-lumi}

  \begin{tabular}[htbp]{rccc}
    \toprule
 \multirow{2}{1cm}{$\mathcal{L}$[fb$^{-1}$]} &
Fraction  with  & Fraction  with    & \multirow{2}{1cm}{$\Delta\sigma/\sigma$}\\
     &
$P(e^{-})=-80\,\%$ &   $P(e^{-})=+80\,\%$  & \\\midrule

    5000     & 50\,\%   & 50\,\%  & 9.0\,\%  \\
    5000     & 80\,\%  & 20\,\% & 7.4\,\%  \\
    5000     & 100\,\% & 0\,\%  & 6.7\,\%  \\
    \bottomrule
  \end{tabular}
\end{table}

Only statistical uncertainties are considered in this study.
Systematic uncertainties for the measurement of the single Higgs
production cross section $\sigma(\Hvv) \times BR(\PH \to \bb)$ from
various potentially dominant sources of systematic uncertainties are evaluated in~\cite{Abramowicz:2210491}.
Potential sources include the luminosity spectrum, the total
luminosity, the beam polarisation, the jet energy scale and flavour tagging.
For the $\sigma(\Hvv) \times BR(\PH
\to \bb)$ measurement, they are shown to be at the per mille level.
As the Higgs bosons in the $\HHvv \to \bbbb$ process are kinematically similar, the systematic uncertainties are expected to be of similar size.
Compared with the almost two orders of magnitude higher statistical uncertainty, the systematic uncertainties are assumed to be irrelevant for this study.

The dependence of the cross section on the value of the trilinear Higgs
self-coupling (Fig.~\ref{fig:xsec_vs_coupling}) is used to derive the projected uncertainty
for the extraction of the trilinear Higgs self-coupling from the
measurement of the cross section.

In order to determine the expected precision for the measurement at
CLIC, a template fit is used based on full detector simulation of
event samples with different values of \gHHH and \gHHWW. A $\chi^2$
minimisation is performed, using the SM sample as the observed
data. For the cases with \gHHWW=\gHHWWSM, pseudo-experiments are drawn
in order to determine the confidence interval at 68\,\% C.L. among the resulting measurements
of \gHHH. 
Based on the measurement of the    \HHvv production
cross section at 3\,TeV only, the expected constraints at 68\,\%
C.L. for \kappaHHH, assuming the SM value for \gHHWW, are $[0.90, 1.12]
\cup [2.40,2.61]$.

\subsection{Precision with \HHvv and \ZHH production at 1.4\,TeV}
\label{sec:ZHH}
One approach that resolves the ambiguity on \gHHH arising from the \HHvv cross-section measurement is the combination with a
measurement of the double Higgsstrahlung cross section, as
described in Sec.~\ref{sec:strategy}.
The estimates were done for $\sqrt{s}$\,=\,1.4\,TeV
 as this is the energy stage of CLIC at which the \ZHH cross section is largest.
 No dedicated full-simulation study has been conducted.
 For illustration, similar analyses have been performed in full simulation for ILC at
$\sqrt{s} = 500$\,GeV and CLIC at $\sqrt{s} = 3$\,TeV.
At  the ILC with $\sqrt{s} = 500$\,GeV~\cite{duerig}, a signal
efficiency of 19\,\% and a background level of 3.6 times the number of
signal events is reached for the hadronic decays of the \PZ boson.
In the CLIC study at 3\,TeV~\cite{Weber:2020hev},
the signal efficiency is 20\,\% with twice the number of background
than signal events.
In both cases, the numbers refer to the analyses with hadronic \PZ
decays and the $\PH \PH \to \bbbb$ channel.

Based on assumptions for different signal efficiencies and background
levels, Table~\ref{tab:ZHH_settings_signficances} lists the
significance with which the Higgsstrahlung process $\PZ\PH\PH$ can
be observed  at $\sqrt{s}=\SI{1.4}{\TeV}$ with
$\mathcal{L}$\,=\,2.5\,ab$^{-1}$ integrated luminosity.
In addition, the CLIC energy stage at 1.4\,TeV will provide a cross-section measurement of the \HHvv process in the \bbbb final state with a luminosity of 2.5\,ab$^{-1}$ and the 4:1 polarisation scheme applied, cf. Table~\ref{tab:HHbbbb_efficiencies_yields_14}.
On its own, this measurement leads to the constraints $[0.64,2.3]$ at
68\,\%\,C.L. in \kappaHHH.
Combining the cross-section measurements of \HHvv and \Zhh at the CLIC
stage at 1.4\,TeV results in the constraints
[0.71, 1.67] for a signal efficiency of 20\,\% and a background level
of twice the signal event number, which is well-motivated by the full-simulation studies described above.
After the CLIC run at the 3\,TeV energy stage, the differential measurement
of \HHvv production will significantly improve those constraints as
described in the next section.
Then, the contributions from the measurements at 1.4\,TeV in \HHvv and
\ZHH will be small (cf. Table~\ref{tab:higgsself-exp-constraints} and Fig.~\ref{fig:chi2_4curves}).

\begin{table}[ht]
  \caption{Significance for \ZHH at 1.4\,TeV in dependence on the assumptions
    for the performance of the ZHH analysis at 1.4\,TeV. The signal
    efficiency $\epsilon_{\text{Sig}}$ and the ratio of selected
    events from background to signal, B/S, are varied.}
  \centering
  \begin{tabular}[ht]{lcc}\toprule
    $\epsilon_{\text{Sig}}$ & B/S &  significance \\\midrule
    50\,\% & 0& 5.7\,$\upsigma$ \\
    40\,\% & 1& 3.6\,$\upsigma$ \\
    30\,\%& 1 & 3.1\,$\upsigma$ \\
    20\,\%& 2 & 2.1\,$\upsigma$ \\
    20\,\%& 3 & 1.8\,$\upsigma$ \\
    14\,\%& 3 & 1.5\,$\upsigma$ \\

    \bottomrule
  \end{tabular}
  
  \label{tab:ZHH_settings_signficances}
\end{table}

\section{Self-coupling extraction based on sensitive kinematic observables}\label{sec:ResultsFromDifferential}
\subsection{Expected precision for the trilinear Higgs self-coupling g$_{\text{HHH}}$}
\begin{figure*}\centering
  
\includegraphics[width=0.87\textwidth]{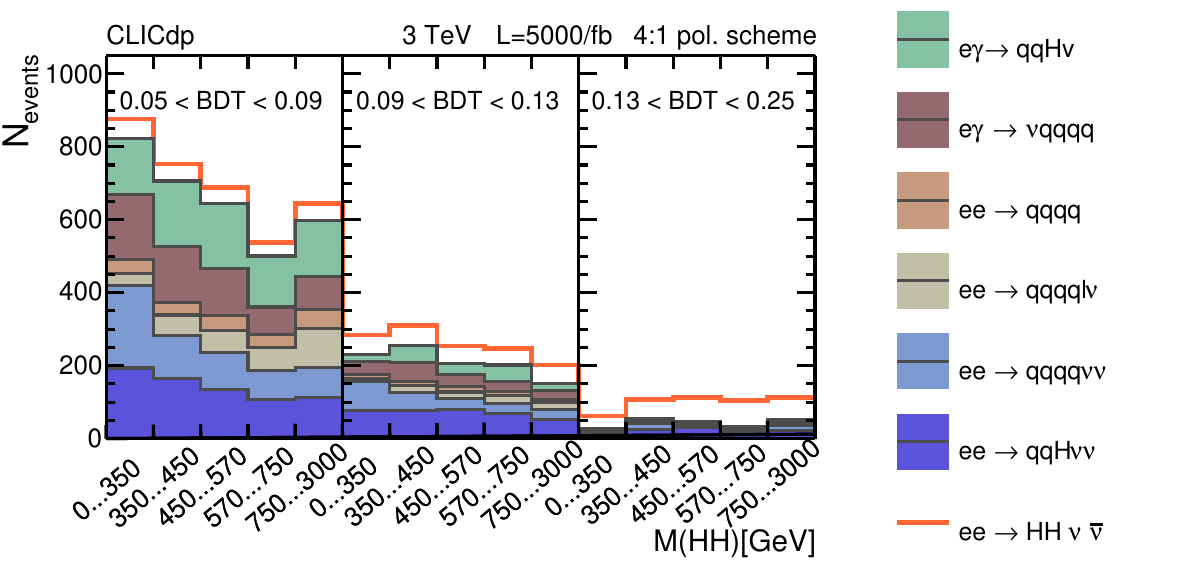}

\caption[Kinematic bins]{Kinematic bins used for the \HHvv sensitivity
  at 3 TeV: the invariant mass of the Higgs boson pair M(HH) in bins of the BDT score.}
\label{fig:BDTvsSumMj_withsignal}
\end{figure*}

Differential distributions sensitive to new physics in the Higgs self-coupling can be used to measure more precisely the trilinear Higgs self-coupling \gHHH and the quartic coupling to W bosons \gHHWW \cite{Contino:2013gna}.
Based on the \bbbb selection, we make use of kinematic observables sensitive to the Higgs self-coupling as described in Sec.~\ref{sec:strategy}.
The highest sensitivity can be reached when using the invariant mass
of the Higgs boson pair in bins of the BDT score.
Fig.~\ref{fig:BDTvsSumMj_withsignal} shows the kinematic bins that are used for a template fit to determine the expected confidence intervals on \gHHH exclusively and on \gHHH and \gHHWW simultaneously.
The one-parameter fit in \gHHH based on the  differential measurement of \HHvv at 3\,TeV
results in expected constraints on \kappaHHH of $[0.92,1.12]$  at
68\,\%\,C.L.

As discussed in Sec.~\ref{sec:ZHH}, the influence of the ZHH measurement
at the second stage is estimated using assumptions based on full
simulation studies.
We therefore assume in the
following that a signal efficiency of 20\,\% and a
background level of twice the signal number can be achieved.
This performance is applied to the
full visible branching fraction of the \PZ boson, although leptonic decay
channels have been found to give much larger signal efficiencies~\cite{duerig}.
Several different cases of signal efficiencies and background levels
are compared in Table~\ref{tab:ZHH_settings_limits} showing that the
resulting uncertainties on the Higgs trilinear self-coupling are
rather stable.

\begin{table}[ht]
  \caption{Constraints on \kappaHHH in dependence on the assumptions
    for the performance of the ZHH analysis at 1.4\,TeV. The signal
    efficiency $\epsilon_{\text{Sig}}$ and the ratio of selected
    events from background to signal, B/S, are varied. To obtain these
    constraints, the \ZHH
    measurement is combined with the full simulation results from the \HHvv
    analyses at 1.4 and 3\,TeV, using  differential information for
    \HHvv at 3\,TeV.}
  \centering
  \begin{tabular}[ht]{lcc}\toprule
    $\epsilon_{\text{Sig}}$ & B/S & ~~~~~ 68\,\% C.L. interval in \kappaHHH \\\midrule
    50\,\% & 0 & [0.92,1.10]\\
    40\,\% & 1 & [0.92,1.10]\\
    30\,\%& 1 & [0.92,1.11]\\
    20\,\%& 2 & [0.92,1.11]\\
    20\,\%& 3 & [0.92,1.11]\\
    14\,\%& 3 & [0.92,1.12]\\

    \bottomrule
  \end{tabular}
  
  \label{tab:ZHH_settings_limits}
\end{table}

\begin{figure}[htbp]
\hspace{-1em} 
   \includegraphics[width=0.53\textwidth]{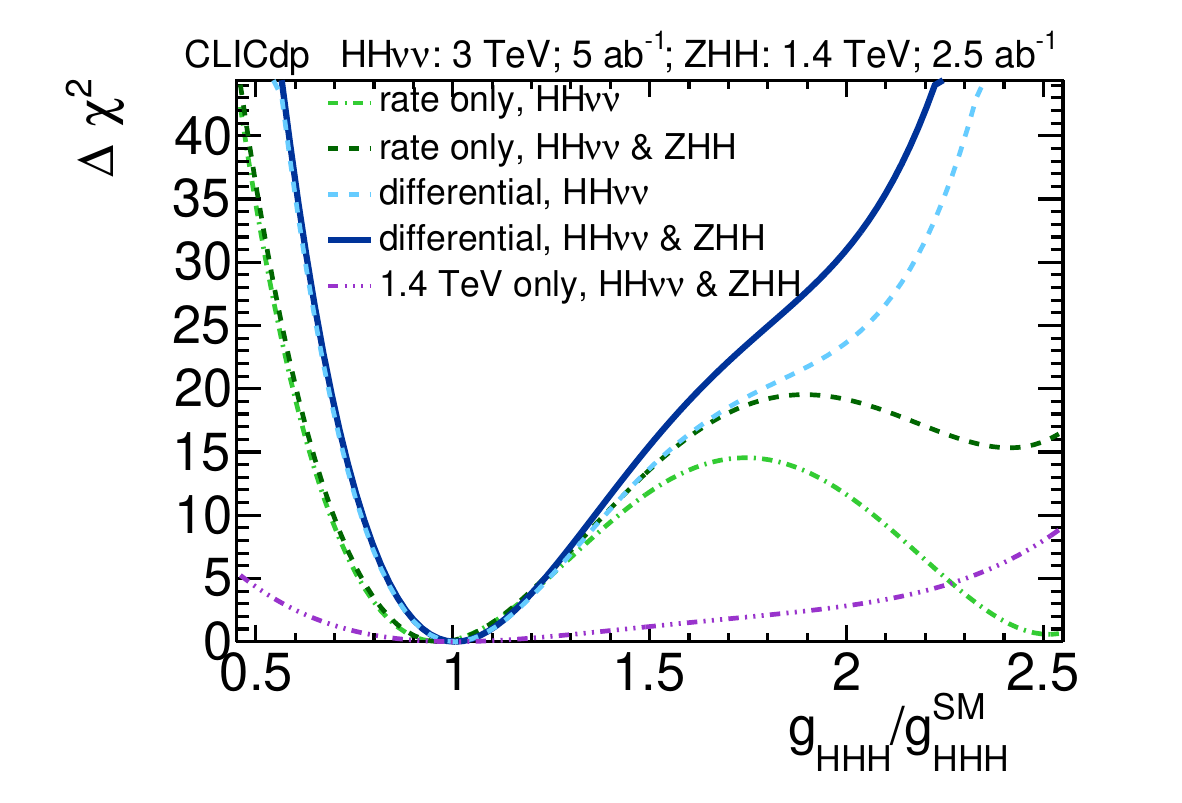}\hspace{1em}\\
 \caption{$\Delta \chi^2$ curves based on rate-only and differential
   information in the \HHvv
   measurement at 3\,TeV  without  and with a combination with the measurement of the
     \ZHH production cross section at 1.4\,TeV. As a comparison, the
     $\Delta \chi^2$ for the case of the second energy stage only is shown.}
   \label{fig:chi2_4curves}
 \end{figure}
 Fig.~\ref{fig:chi2_4curves} illustrates the resulting
 \deltachisquared curves from the different steps of the analysis:
 Adding the information from the \ZHH analysis to the rate-only
 measurement of \HHvv raises the second minimum above the 68\,\%~C.L.
However, with the differential measurement at 3\,TeV alone, the second
minimum is removed already, and the  expected constraints for \kappaHHH at
68\,\%\,C.L. are $[0.92,1.12]$, as discussed above.
In this case, the impact of the \ZHH analysis at 1.4\,TeV is small.
By combining the differential
analysis of \HHvv with the \ZHH and \HHvv cross-section measurements
at 1.4\,TeV, the best constraints are obtained, reaching $[0.92,1.11]$
at 68\,\%\,C.L.
This is the final resulting expectation for the sensitivity of the full CLIC programme to the trilinear Higgs self-coupling using the invariant di-Higgs mass and the BDT score as template.
Table~\ref{tab:higgsself-exp-constraints} summarises the 68\,\%\,C.L. constraints obtained for \gHHH/\gHHHSM with the different approaches.

\begin{table*}
  
  \caption[Constraints on \kappaHHH]{Constraints on
    \kappaHHH obtained in the full detector simulation
    study using a multivariate analysis for selection. The constraint
    from cross section only is obtained in the tight BDT selection. The constraints based on differential
    distributions are derived in the loose BDT selection.}
\label{tab:higgsself-exp-constraints}
  \centering
\begin{tabular*}{\linewidth}{@{\extracolsep{\fill}}*1l@{}@{\extracolsep{\fill}}*1c@{}}
    \toprule
 Constraints for $ \kappa_{\HHH}$ based on  & $\Delta \chi^{2 } =1 $\\ \midrule
  \HHvv cross section only (3\,TeV)                                                             & $[0.90, 1.12]$ $\cup$ $[2.40,2.61]$ \\
 \HHvv differential (3\,TeV)                                                                   & $[0.92, 1.12]$                      \\
 \HHvv differential (3\,TeV) and  \ZHH  (1.4\,TeV) cross section                               & $[0.92, 1.12]$                      \\
 \HHvv differential (3\,TeV), \HHvv cross section (1.4\,TeV) and \ZHH cross section (1.4\,TeV) & $[0.92, 1.11]$                      \\
\bottomrule
  \end{tabular*}\end{table*}

These results can be interpreted in scenarios of new physics
modifying the Higgs self-coupling.
An example is the case of a Higgs plus singlet model~\cite[Sec.~6.1]{BSRYR} and \cite{No:2018fev}, where  a general real singlet scalar is added to the SM Higgs sector. This could lead to a strong first-order electroweak phase transition and therefore offers an explanation of baryogenesis.
The model introduces a heavy singlet as well as a mass eigenstate mixing with the SM-like Higgs boson. 
The new parameters are therefore the mass of the heavy scalar and the
mixing angle between the singlet and the SM-like Higgs boson, as well as the parameters of the scalar potential. 
The parameter space is constraint by theoretical considerations such as unitarity and perturbativity. In addition, the EW vacuum is required to be stable.
Considering direct searches in resonant double Higgs boson production
as well as the sensitivity to the trilinear Higgs self-coupling, CLIC
is sensitive to a sizable fraction of the parameter space which is also compatible with a strong first-order EW phase transition.
In most cases, single Higgs boson coupling measurements give a similar reach as the Higgs boson pair searches, allowing a complementary assessment of the implications on the electroweak phase transition.

\subsection{Expected precision for simultaneous fit of g$_{\text{HHH}}$ and  g\boldmath{$_{\text{HHWW}}$}}

As described in Sec.~\ref{sec:strategy}, the Higgs-gauge vertex \HHWW contributes to \HHvv as well.
We can therefore extend the study of \HHvv production at 3\,TeV 
to fit simultaneously the modified couplings \kappaHHH and \kappaHHWW.
All other EFT couplings are kept at the SM value, in particular the coupling $g_{\PZ\PZ\PH\PH}$.
Based on the 
differential distribution and binning depicted in
Fig.~\ref{fig:BDTvsSumMj_withsignal}, we determine the 68\,\% and
95\,\%\,C.L. contours for two degrees of freedom.
The deviation of the nominal samples from the SM by $\Delta
\chi^2=2.3$ ($6.18$) is used for the constraints at 68\,\% (95\,\%) C.L. 

\begin{figure}[htbp]\centering
    \includegraphics[width=0.49\textwidth]{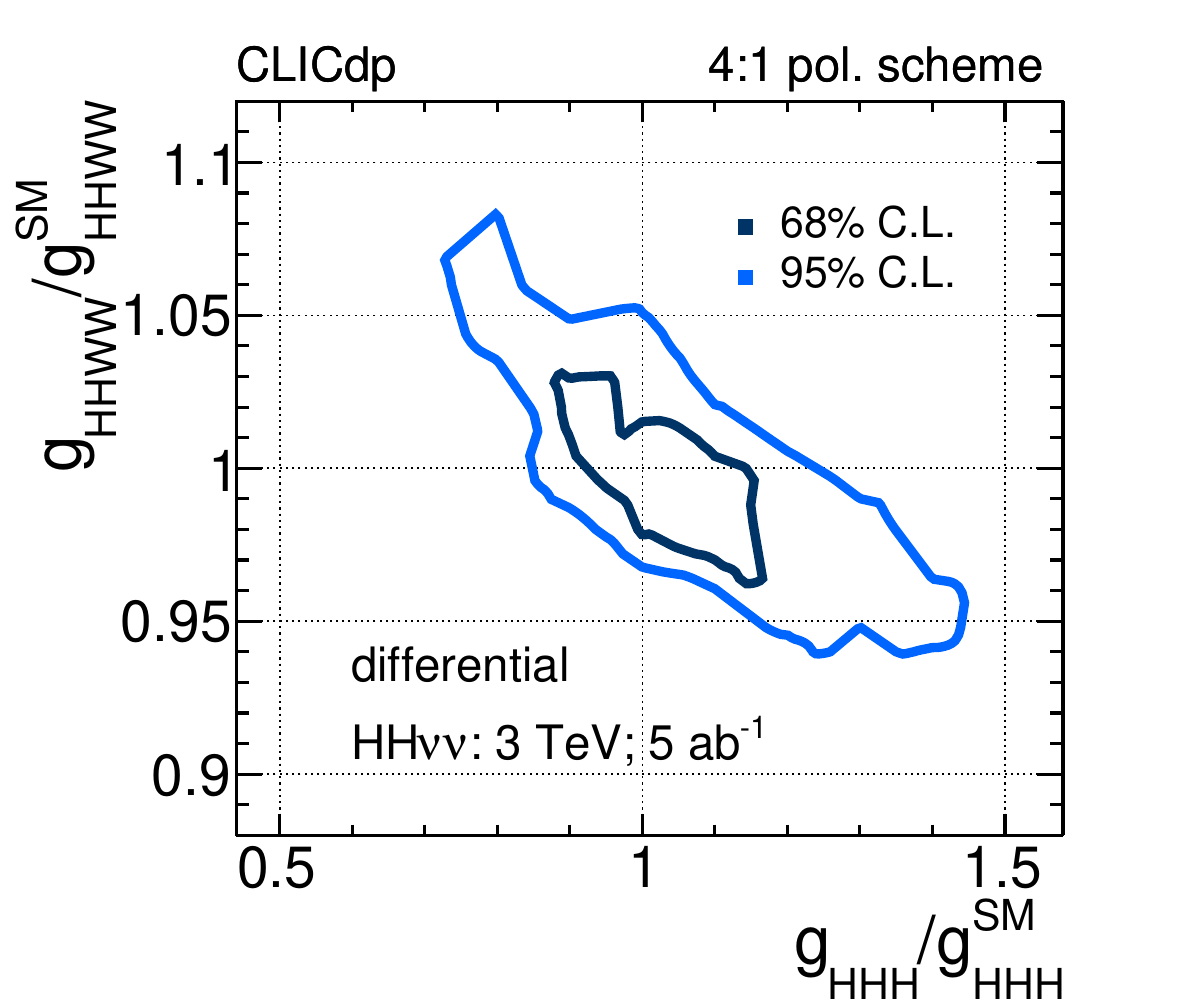}
  \caption{Confidence contours at 68\,\% and 95\,\%\,C.L. for the
    simultaneous fit of \kappaHHH and \kappaHHWW based on differential
    measurement in \HHvv production at 3\,TeV CLIC.} 
\label{fig:contour2D}
\end{figure}

The resulting constraints are shown in Fig.~\ref{fig:contour2D}. At 68\,\% C.L. the
simultaneous fit leads to expected constraints of up to $20\,\%$ in
\kappaHHH   and up to  4\,\% in \kappaHHWW across the allowed range of the other coupling.
Due to the anticorrelation illustrated in Fig.~\ref{fig:contour2D}, the individual constraints for fixed values of the other coupling are substantially smaller.
Going beyond the two effective couplings \gHHWW and \gHHH, this measurement can be combined with other measurements at CLIC in
order to perform a global EFT fit of the full set of relevant operators.

\section{Conclusions}\label{sec:summary}

In this paper, the prospects for the extraction of the trilinear Higgs self-coupling and the quartic \HHWW coupling at CLIC are presented.
The results are based on double Higgs-boson production in the processes $\epem\to$ \HHvv and $\epem \to $ \ZHH.
Analyses of \HHvv production have been performed for the decay channels \bbbb and \bbWW in full simulation.
The analyses assume the second and third stage of CLIC at collision energies of 1.4\,TeV and 3\,TeV.
In addition, the contribution of the \ZHH cross-section measurement has been included for 1.4\,TeV.
The channel with the highest sensitivity to the Higgs self-coupling and the \HHWW coupling at CLIC is the \bbbb decay channel of \HHvv production at 3\,TeV, where the total cross-section measurement as well as differential distributions can be used to extract the couplings.
The differential measurement is based on the invariant mass distribution of the double Higgs-boson system as well as a multivariate score.

Generally, this channel can be useful to study the impact of heavy flavour tagging  and jet energy resolution, especially in the forward direction, realised in the CLIC detector models. In this case, the CLIC\_ILD model was used. No significant change is expected for the application of this analysis to the current CLICdet model.
This analysis benefits from the higher centre-of-mass energy due to the increase in cross section of \HHvv production.
It therefore provides a strong motivation for the CLIC 3\,TeV energy stage.

Beams in the simulated samples used in
  this analysis are unpolarised. Based on the cross section for each polarisation mode, the results are scaled to the baseline
  polarisation scheme of CLIC with the running time shared between
  P(\Pem)\,=\,$-80\,\%~(+80\,\%)$ in the ratio 4:1. Future studies making use of the
  polarisation-dependent kinematic behavior might improve the signal
  selection.
Furthermore, future studies could
      treat the \ZvvHH component separately in the \HHvv final state. This contribution is particularly important at
      1.4\,TeV and also depends on the polarisation mode. As it also
      has a dependence on \gHHH, albeit a different one than the WBF
      production channel, this can be exploited separately. 

At the 1.4\,TeV energy stage of CLIC, evidence for the \HHvv process of the SM can be reached with a significance of 3.5\,$\upsigma$. With a luminosity of only 700\,fb$^{-1}$, the process can be observed with 5.0\,$\upsigma$ at 3\,TeV.
Taking into account only the 1.4\,TeV stage of CLIC with cross-section measurements of \HHvv and \Zhh allows the measurement of the Higgs self-coupling \gHHH with relative uncertainties of $-29\,\%$ and $+67\,\%$ around the SM value at 68\,\%\,C.L.
Based on events of double Higgs-boson production at both high-energy stages, CLIC can be expected to
 measure the trilinear Higgs self-coupling \gHHH with a relative uncertainty of $-8\,\%$ and $ +11\,\%$ at 68\,\% C.L., assuming the Standard Model and setting the quartic \HHWW coupling to its Standard Model value.
Measuring simultaneously the trilinear Higgs self-coupling and the quartic Higgs-gauge coupling results in constraints at 68\,C.L. below 4\,\% in \gHHWW and below 20\,\% in \gHHH for large modifications of \gHHWW.
These results illustrate the strength of the proposed CLIC programme to make a precise measurement of the trilinear Higgs self-coupling.

\begin{acknowledgements}
  This work benefited from services provided by the ILC Virtual Organisation, supported by the national resource providers of the EGI Federation.
This research was done using resources provided by the Open Science Grid, which is supported by the National Science Foundation and the U.S. Department of Energy's Office of Science.

\end{acknowledgements}

\bibliographystyle{spphys}       
\bibliography{bibliography}   

\end{document}